\documentclass[preprint,authoryear,12pt]{elsarticle}
\usepackage{amssymb}
\journal{Advances in Space Research}
\usepackage[columnwise]{lineno}
\usepackage{lipsum}
\usepackage{multicol}
\usepackage{blindtext}
\usepackage[pdfborder={0 0 0 },urlcolor=blue]{hyperref}
\ifx \doiurl \undefined \def \doiurl#1{\href{http://dx.doi.org/#1}{\url{#1}}}\fi 
\ifx \adsurl \undefined \def \adsurl#1{\href{http://adsabs.harvard.edu/abs/#1}{\url{#1}}}\fi

\newcommand{\etal}{{\it et al.}}
\newcommand{\adv}{    {\it Adv. Space Res.}}

\newcommand{\aap}{    {\it Astron. Astrophys.}}

\newcommand{\apj}{    {\it Astrophys. J.}}
\newcommand{\apjl}{   {\it Astrophys. J. Lett.}}

\newcommand{\apss}{   {\it Astrophys. Space Sci.}}

\newcommand{\mnras}{  {\it Mon. Not. Roy. Astron. Soc.}}
\newcommand{\nat}{    {\it Nature}}

\newcommand{\solphys}{{\it Solar Phys.}}

\newcommand{\Natco}{    {\it Nat. Commun.}}
\newcommand{\na}{    {\it NewA}}
\chardef\us=`\_

\begin{document}
\begin{frontmatter}

\title{Prediction of the amplitude of solar cycle 25 from  
the ratio of sunspot number to sunspot-group area, 
 low latitude activity, and 130-year solar cycle}

\author{J.\ Javaraiah}

\address{Bikasipura, Bengaluru-560 111,  India.\\
Formerly working at Indian Institute of Astrophysics, Bengaluru-560 034,
India.
}
\ead{jajj55@yahoo.co.in;  jdotjavaraiah@gmail.com; jj@iiap.res.in}

\begin{abstract}
Prediction of the amplitude of  solar cycle is important for 
understanding the mechanism of solar cycle and solar activity 
influence on space-weather. 
We analysed the combined 
data of sunspot groups from Greenwich Photoheliographic Results (GPR) 
during the period 1874\,--\,1976 and 
Debrecen Photoheliographic Data (DPD) during 1977\,--\,2017 and 
determined the monthly mean, annual mean, and 13-month smoothed
 monthly mean whole sphere  sunspot-group area
 (WSGA). We also analysed the  monthly mean, annual mean, and  
  13-month smoothed 
 monthly mean version 2 of international sunspot number ($SN_{\rm T}$)
 during the period 1874\,--\,2017.  We fitted the 
  annual mean  WSGA and $SN_{\rm T}$ data during each of
 Solar Cycles~12\,--\,24 separately to the linear and nonlinear 
(parabola)
 forms. In the cases of Solar Cycles 14, 17, and 24 the nonlinear fits 
are found better than the linear fits.
  We find that there exists a secular
 decreasing trend in the slope   of the
 WSGA--$SN_{\rm T}$ linear relation during Solar Cycles~12\,--\,24.  A secular 
decreasing trend is also seen in the coefficient of the first order term 
of the nonlinear relation. 
 The existence of $\approx$77-year variation is  clearly seen 
in the ratio of the amplitude to WSGA  at the maximum
 epoch of solar cycle. From  the pattern of this long-term variation of the 
 ratio we inferred that Solar Cycle~25  will be larger than 
both Solar Cycles 24 and 26. Using an  our earlier 
method (now slightly revised), i.e. using   
 high correlations of the amplitude of a solar cycle with the sums
 of the areas of sunspot groups in $0$\,--\,$10^\circ$ latitude intervals
 of the northern hemisphere during 3.75-year interval around the 
minimum--and the  southern hemisphere during  0.4-year interval
 near the maximum--of the corresponding preceding solar cycle,
 we predicted   
 $127 \pm 26$ and $141 \pm 19$ for the amplitude of Solar Cycle~25, 
respectively.
Based on $\approx$130-year periodicity found in the cycle-to-cycle
variation of the amplitudes of Solar Cycles 12\,--\,24 we find the shape of
 Solar Cycle~25 would be similar to that of Solar Cycle~13 and predicted  
 for Solar Cycle 25 the amplitude $135 \pm 8$, maximum epoch
 2024.21 (March 2024)$\pm$6-month, and the following minimum epoch  
   2032.21 (March 2032)$\pm$6-month with $SN_{\rm T}$ $\approx 4$.
\end{abstract}


\end{frontmatter}
\parindent=0.5 cm

\section{Introduction}
Prediction of the amplitude of solar cycle is very important because 
it helps  for understanding the mechanism of solar cycle,  solar activity
influence on space weather, and solar-Terrestrial relationship \citep{hath15}.
 Many 
techniques are used for predicting the amplitudes of Solar Cycles~24 
and 25 \citep{pes12,petro20,nandy21}.
The sunspot number and sunspot area are the most prominently used indece of 
solar activity to study characteristics of solar cycles and to investigate 
other short- and long-term variations of solar activity. It is well-known
 that there exists a good linear 
relationship between the maximum sunspot number and the maximum sunspot area 
of solar cycles \citep[e.g.,][]{hath02,jj22}. However, the 
characteristics of sunspot-number cycles and sunspot-area cycles  are not 
exactly the same.  In some solar cycles the epochs of maxima of sunspot
 number and sunspot area are to some extent different \citep{ramesh08}. 
The ratio of monthly or yearly mean  sunspot area to  
sunspot number varies during a solar cycle, it is large during maximum
and least during minimum of a solar cycle. However, in many solar cycles there
 exist considerable differences in the positions of the maximum and minimum of 
the ratio with respect to the corresponding sunspot number maximum and minimum
 \citep{wh06}.
The well-known Waldmeier effect of sunspot cycles
 (large cycles rise faster than the small cycles)  seems not apply for
 sunspot-area cycles  \citep{dgt08,jj19}.  However,  
 \cite{kc11} by correcting the positions of the area-peaks of some solar cycles
  obtained an anticorrelation between rise time and  amplitude. There 
exits a good correlation between  rise rate and 
 amplitude \citep{kc11,kumar22}.
  Large and small sunspots/sunspot-groups are not in the same propositions  
 in all solar cycles and there exist differences in their 
variations during solar cycles \citep{kilc11,jj12,jj21,clette16,mandal16}. 
 Moreover, sunspot areas are thought to be 
more physical measures of solar activity  (Hathaway, 2015).  It is
 believed that  sunspot area represents the  solar magnetic flux better
 than sunspot number.

 In this analysis  we determined the correlation between  
the annual mean values of total (north $+$ south) sunspot number 
$SN_{\rm T}$ and the annual mean area (WSGA) of the
 sunspot groups in whole sphere (north $+$ south)  during  each of 
Solar Cycles 12\,--\,24, separately.
 We fitted the
  annual mean values of WSGA and $SN_{\rm T}$  during each of
 Solar Cycles~12\,--\,24 separately to the linear and nonlinear
 (parabola) 
 forms.  In a few solar cycles the fit of  
 the nonlinear form is found better than the linear form.
  We show that  there exists a secular
 decreasing trend in the slope   of the
 WSGA--$SN_{\rm T}$ linear relation during Solar Cycles~12\,--\,24.
 We also show 
that there exists a secular
decreasing trend mainly in the coefficient of the first order term
of the nonlinear relation.
 In this analysis we also determined the  
 solar cycle-to-cycle  variation in the ratio of the 
  smoothed monthly mean maximum value of  
  $SN_{\rm T}$ (i.e. the amplitude $R_{\rm M}$) of a solar cycle to the
 smoothed monthly mean value of WSGA at the maximum epoch of the solar cycle.
Based on the pattern of this variation 
we make predictions for the relative amplitudes of some upcoming solar 
cycles.  By making a minor change in  our earlier
 method that based on low latitude activity \citep{jj07,jj08} we have   
 made improved predictions for the amplitude of Solar Cycle 25. 
Recently, \citep{jj22,jj23}, we have noticed  
 that there  exists a $\approx$130-year periodicity in the 
amplitude ($R_{\rm M}$) modulation during  Solar Cycles~12\,--\,24.
 The uncertainty in this periodicity is to some extent large because 
it is determined from only 13 sunspot cycles' data. However, this periodicity 
is found in the huge data of solar activity related phenomena, 
such as  $^{14}C$ and $^{10}Be$ \citep{att90a,att90b,mcc13}.
 On  the  basis of this periodicity  we 
find that the shape of  Solar Cycle 24 is similar to that of Solar Cycle~12 
and the rising  phase of Solar Cycle~25 is similar to that of  Solar Cycle~13. 
  Using it we predict the shape of Solar Cycle 25 will be similar to 
that of Solar
 Cycle~13 and   predict the amplitude, maximum epoch, and the ending
 epoch of Solar Cycle~25.   

In the next section we describe the  data and analysis. In Section~3 
we present the results. In Section~4 we present the conclusions and
discuss them.

\section{Data and analysis} 
We have used the data of sunspot groups  from Greenwich Photoheliographic 
Results (GPR) during the period 1874\,--\,1976 and Debrecen Photoheliographic 
Data (DPD) during the period  1977\,--\,2017. These data are downloaded from 
the  website {\sf fenyi.solarobs.unideb.hu/pub/DPD/}.  These data 
  contain heliographic positions (latitude and 
longitude), the corrected whole-spot area ($A$), 
central meridian distance (CMD),
 etc. for each sunspot group for each day  during its lifetime (disk passage).
In order to reduce the foreshortening effect (if any) we have used only
 the values  of $A$ correspond to the $|{\rm CMD}| \le 75^\circ$.   
 We determined the mean  
  whole sphere sunspot-group area (WSGA) of each month during the period 
1874\,--\,2017 and then we obtained the time series of 13-month smoothed
 monthly mean   and  annual mean values of WSGA.  
  We have used  the time series of 13-month smoothed
 monthly mean values, monthly mean values, and  annual values of
 version-2 of the total international sunspot number ($SN_{\rm T}$) 
during the period 1874\,--\,2023 (downloaded the files SN\_ms\_tot\_v2.0.txt, 
 SN\_m\_tot\_v2.0.txt,  and SN\_y\_tot\_v2.0.txt
  from the website {\sf www.sidc.be/silso/datafiles}). The subscript `T' 
of $SN_{\rm T}$ indicates northern and southern hemispheres' total.   
 All these time series are partitioned for individual
 Solar Cycles 12\,--\,24.  
 We fitted each solar cycle's annual mean values (also monthly and smoothed 
monthly values) of  WSGA and $SN_{\rm T}$ 
  to the linear  from: 
\begin{equation}
SN_{\rm T} = m \times {\rm WSGA}, 
\label{eq1}
\end{equation}
\noindent and also to the nonlinear from: 
\begin{equation}
SN_{\rm T} = m_1 \times {\rm WSGA} + m_2\times {\rm WSGA}^2.
\label{eq2}
\end{equation}
Both these  equations are forced to pass through the origin by assuming
 $SN_{\rm T}$ is zero when WSGA is zero.  The linear least-square fits
  are calculated by using the Interactive Digital Library (IDL) software
 {\textsf{FITEXY.PRO}}, which  is downloaded from
 the website \textsf{http:\ //idlastro.gsfc.nasa.gov/ftp/pro/math/}.
An advantage of this software is it takes care the uncertainties in 
 the values of both the abscissa (WSGA) and ordinate ($SN_{\rm T}$) in 
the calculations of linear-least-square fit.  
The nonlinear fit is done by taking into account 
  only the  uncertainties in $SN_{\rm T}$ (in the available 
software there is no provision for using the errors in the abscissa). 
 We also determined the
 solar cycle-to-cycle  variation of the ratio $R_{\rm M}/R_A$ during Solar
 Cycles 12\,--\,24,  where the amplitude $R_{\rm M}$ and $R_A$ are the
 13-month smoothed monthly mean values of $SN_{\rm T}$ and WSGA, respectively,
 at the maximum epoch of a solar cycle  (this is also done by using the 
 corresponding annual mean values).

 In our earlier papers, \citep{jj15,jj17,jj21,jj23}, based on
an high correlation  was found  between  the amplitude ($R_{\rm M}$) of 
a solar cycle
and the sum of the areas ($A^*_{\rm M}$) of the sunspot
 groups in  $0$\,--\,$10^\circ$ latitude interval of the southern
 hemisphere during a  slightly less than one-year time interval ($T^*_{\rm M}$)
 just after the maximum of the solar cycle (namely REL-I),
we have predicted that the
 amplitude of Solar Cycle 25 will be considerably weaker than the amplitude of
 the reasonably small Solar Cycle 24. That prediction is incorrect   
 (the height of the rising phase of Solar Cycle 25 is already higher than
 that  of Solar Cycle 24).
 On the other hand, all our earlier analyses suggest
 a low amplitude for Solar Cycle~25. As per the current trend of activity
this suggestion is expected  to be holds good, although this cycle will be
slightly larger than Solar Cycle~24. In addition, based on
 the pattern of the cycle-to-cycle variation in the aforementioned
sum of the areas of the sunspot groups  we also predicted 
 as Solar Cycle~25 will be larger than Solar Cycle~24
\citep{jj08}, which is consistent with the current trend of the rising phase 
of Solar Cycle~25.

Earlier, \citep{jj07}, we have also found an
 high correlation   between  the amplitude of a solar cycle
and the sum of the areas ($A^*_{\rm m}$) of sunspot
 groups in  $0$\,--\,$10^\circ$ latitude interval of the northern
 hemisphere during about 3.5-year time interval ($T^*_{\rm m}$)
 around the preceding
minimum of  previous solar cycle (namely REL-II).
However, the corresponding correlation of this relationship (REL-II)
 is relatively smaller than that was obtained by using REL-I
and also the uncertainty in the predicted value of the amplitude of
Solar Cycle~24 was found relatively large. Later the predicted
 amplitude of Solar Cycle~24 was found considerably larger than the
 observed value.
The corresponding prediction made by using REL-I was found
closely match with the observed value. Therefore, in \cite{jj15,jj17}
 we have not used REL-II for predicting the amplitude of Solar Cycle~25.
 In \cite{jj21} we have used REL-II and obtained $123\pm 23$ for the amplitude
of Solar Cycle~25, but this prediction was not claimed there.

Here we analysed the GPR and DPD sunspot group data during
1874\,--\,2017 and  as in  \citet{jj07,jj08}  we determine
the relations  REL-I and REL-II by taking into account 
the uncertainties in
the values of the maxima of Solar Cycles 12\,--\,24 and making corrections 
to the values of $T^*_{\rm M}$ of Solar Cycles 23 and 24. 
Using the improved relations REL-I and REL-II  we make improvements in
 our earlier  predictions of the amplitude  of Solar Cycle~25.  

Based on the existence of a $\approx$12-cycle ($\approx$130-year) periodicity 
in the variation of a $R_{\rm M}$ during Solar Cycles 12\,--\,24 
\citep{jj22,jj23} we found 
that the shape of Solar Cycle~24 is similar to that of Solar Cycle~12 and the  
rising phase of Solar Cycle~25 (for which the values of $SN_{\rm T}$ are 
available) is similar to that of Solar Cycle~13. We find a high 
correlation between the variation in the 13-month smoothed monthly mean 
$SN_{\rm T}$  during 
Solar Cycles~12\,--\,13 (rise phase) and that during 
Solar Cycles~24\,--\,25 (rise phase). 
The corresponding linear-least-square best-fit is used to simulate the 
 Solar Cycles~25\,--\,26.

\section{Results}
\subsection{Relationship between sunspot number and sunspot-group area} 
\begin{figure}
\centering
\includegraphics[width=\textwidth]{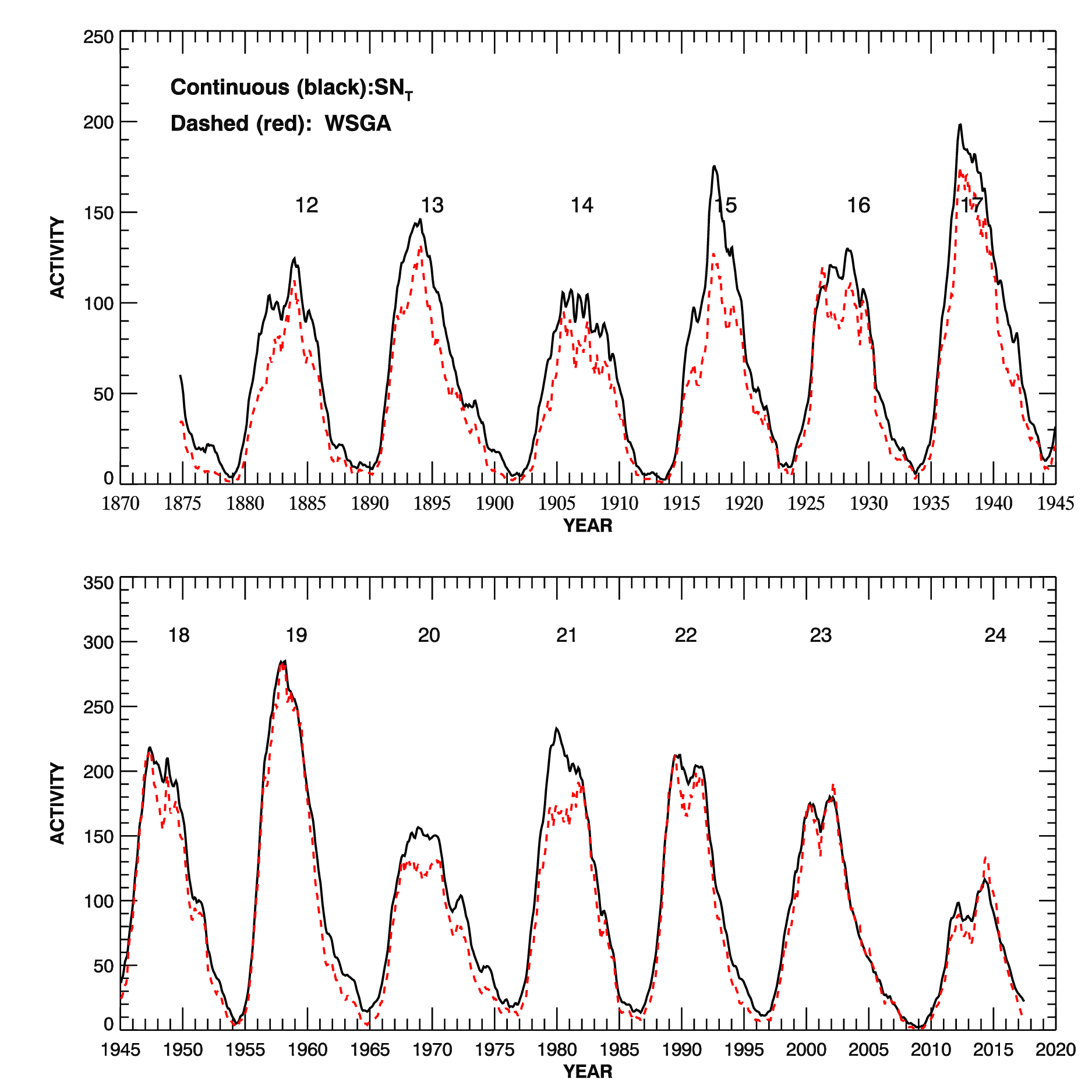}
\caption{Variations in the 13-month smoothed monthly mean version 2 of  
international sunspot number $SN_{\rm T}$  (black {\it continuous-curve})
 and the corresponding smoothed  
area of the  sunspot groups in the whole sphere WSGA (red {\it dashed-curve}) 
during the period 1874\,--\,2017. The values of WSGA is first divided by the
 largest value of WSGA, 3480.15 msh, and then multiplied by the largest value,
 285.0 of $SN_{\rm T}$. Waldmeier numbers of the solar cycles are also shown.}
\label{f1}
\end{figure}

Fig.~\ref{f1} shows variations in the 13-month smoothed monthly mean version~2
 of
international sunspot number ($SN_{\rm T}$) and the corresponding smoothed
area (WSGA) of the sunspot groups in the whole sphere 
during the period 1874\,--\,2017. As can be seen in this figure there are 
some noticeable differences in the variations of $SN_{\rm T}$ and WSGA, 
during many solar cycles  \citep[also see][]{jj19,jj20}.  A good 
 agreement between the variations of $SN_{\rm T}$ and WSGA
 during the minima of most of 
the solar cycles, whereas some considerable differences exist during the maxima of many solar cycles. 
That is, although  the positions of the peaks during the  
maxima (i.e., Gnevyshev peaks) of $SN_{\rm T}$-cycles and WSGA-cycles 
are almost the same, but in some cycles (e.g., 16 and 21) the main (highest)
 and second high (secondary) peaks are interchanged. In the case of some 
solar cycles
 (16, 17, 20, 21) the heights of the peaks of WSGA are relatively lower than
 those of $SN_{\rm T}$, and it seems opposite in Solar Cycles~23 and 24 
 (see the main peaks). 

\begin{figure}
\centering
\includegraphics[width=3.5cm]{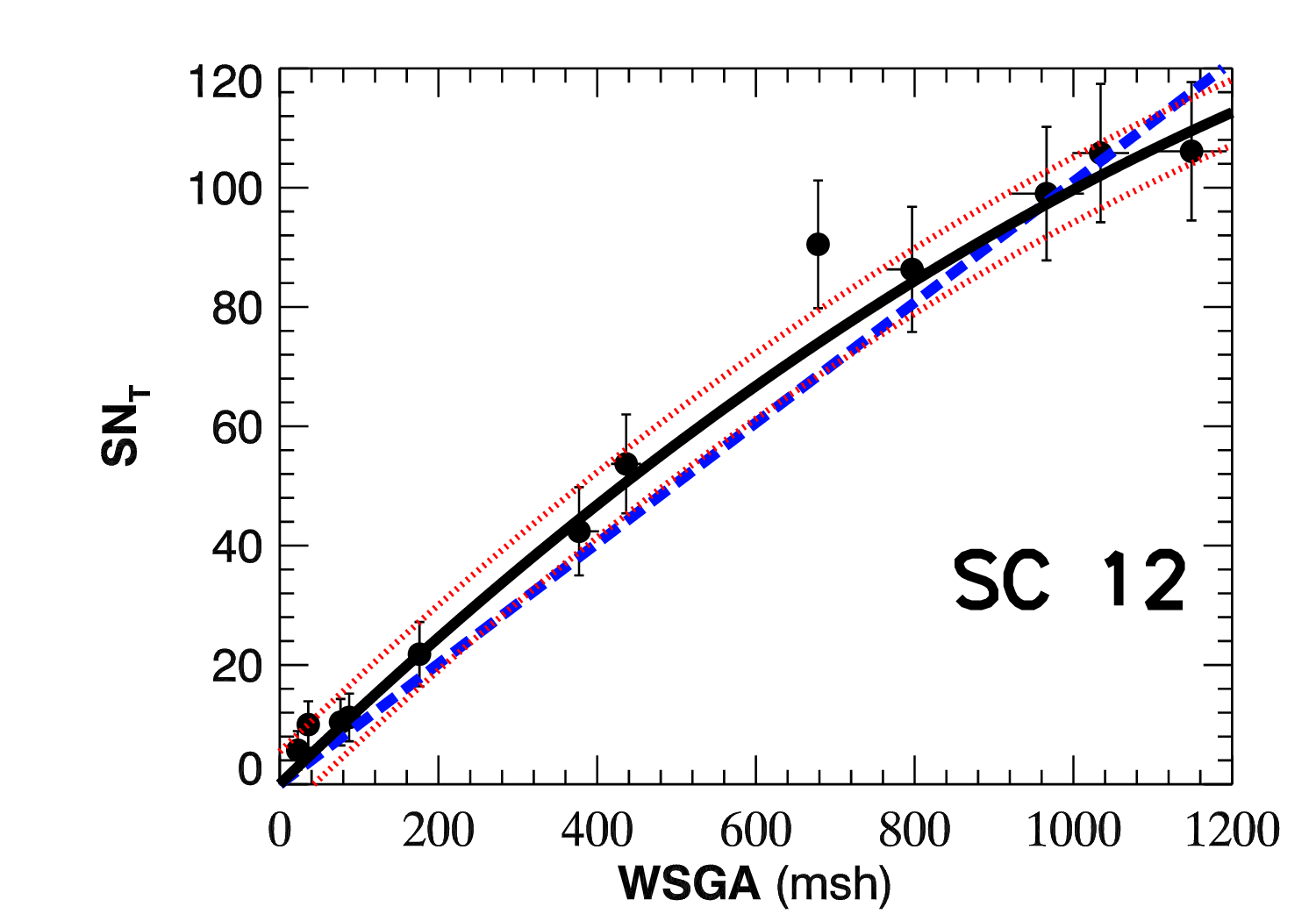}
\includegraphics[width=3.5cm]{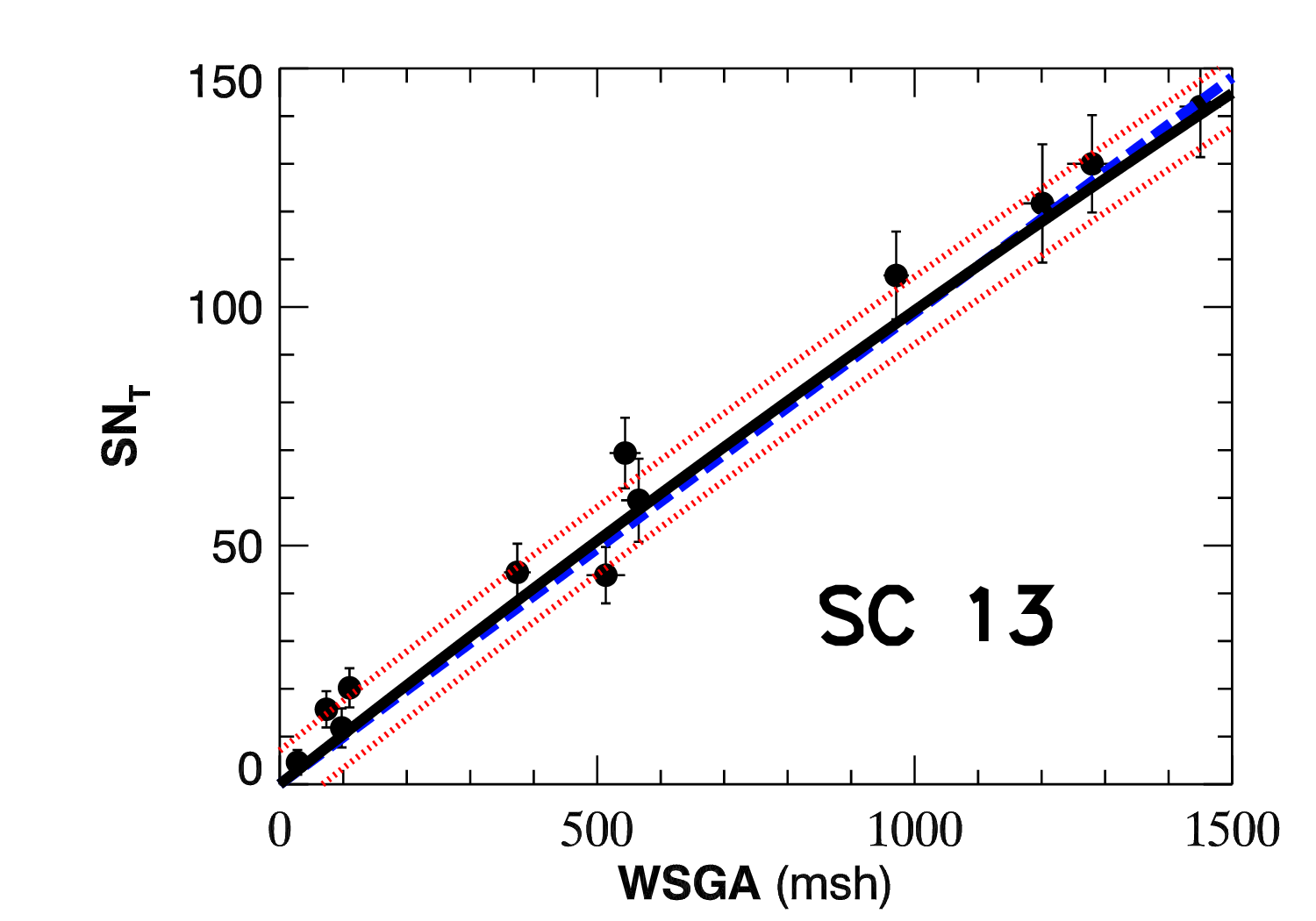}
\includegraphics[width=3.5cm]{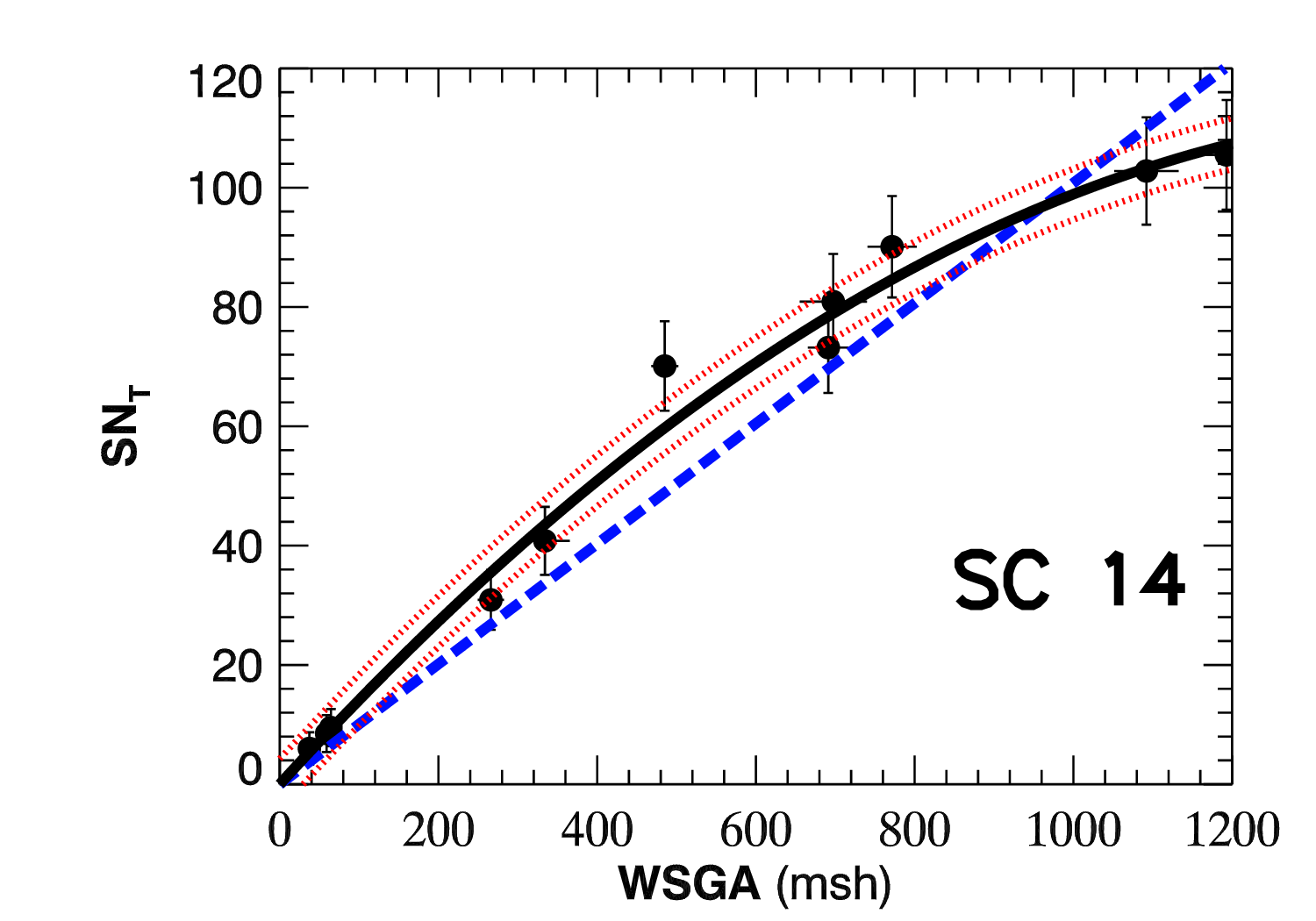}
\includegraphics[width=3.5cm]{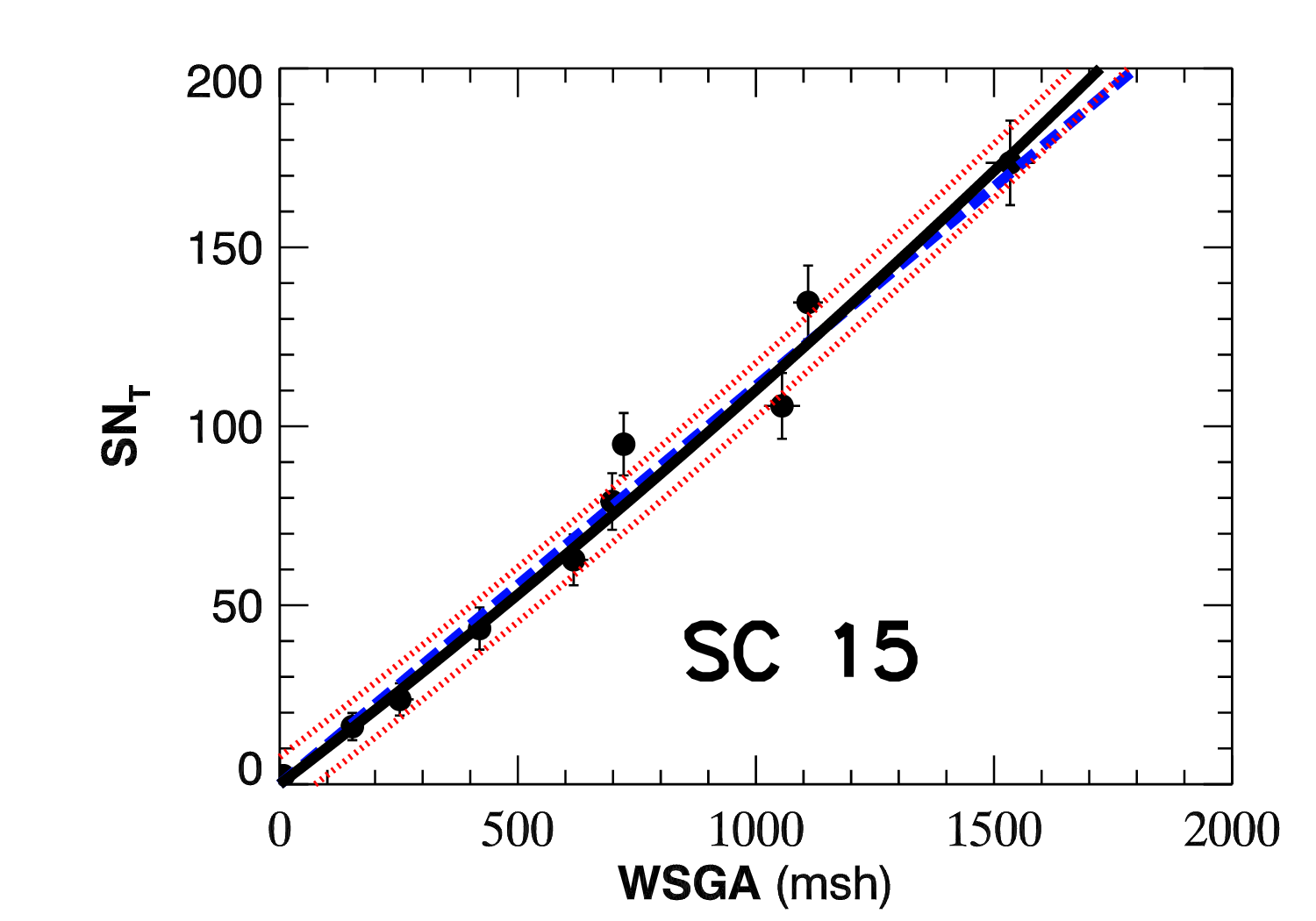}
\includegraphics[width=3.5cm]{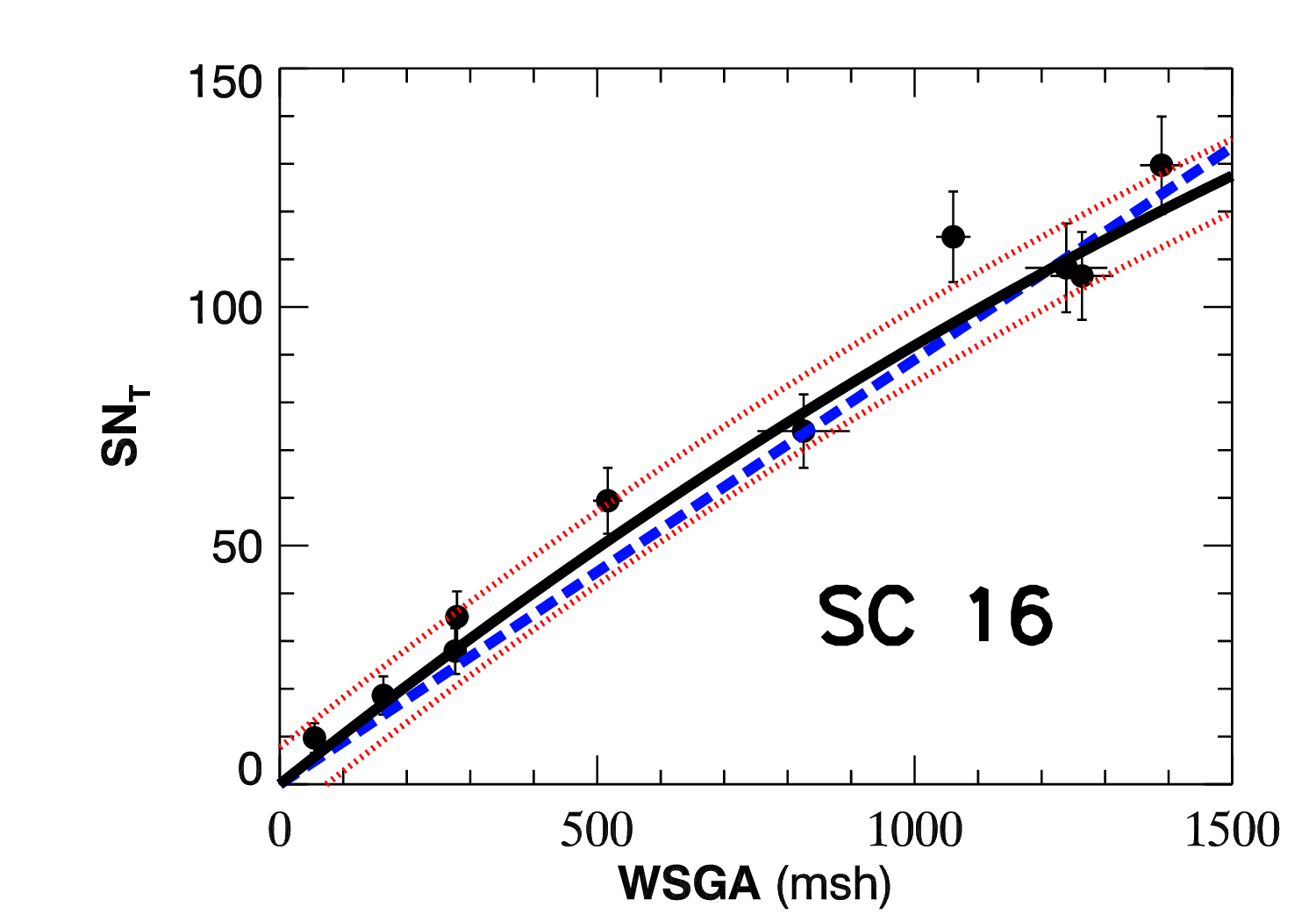}
\includegraphics[width=3.5cm]{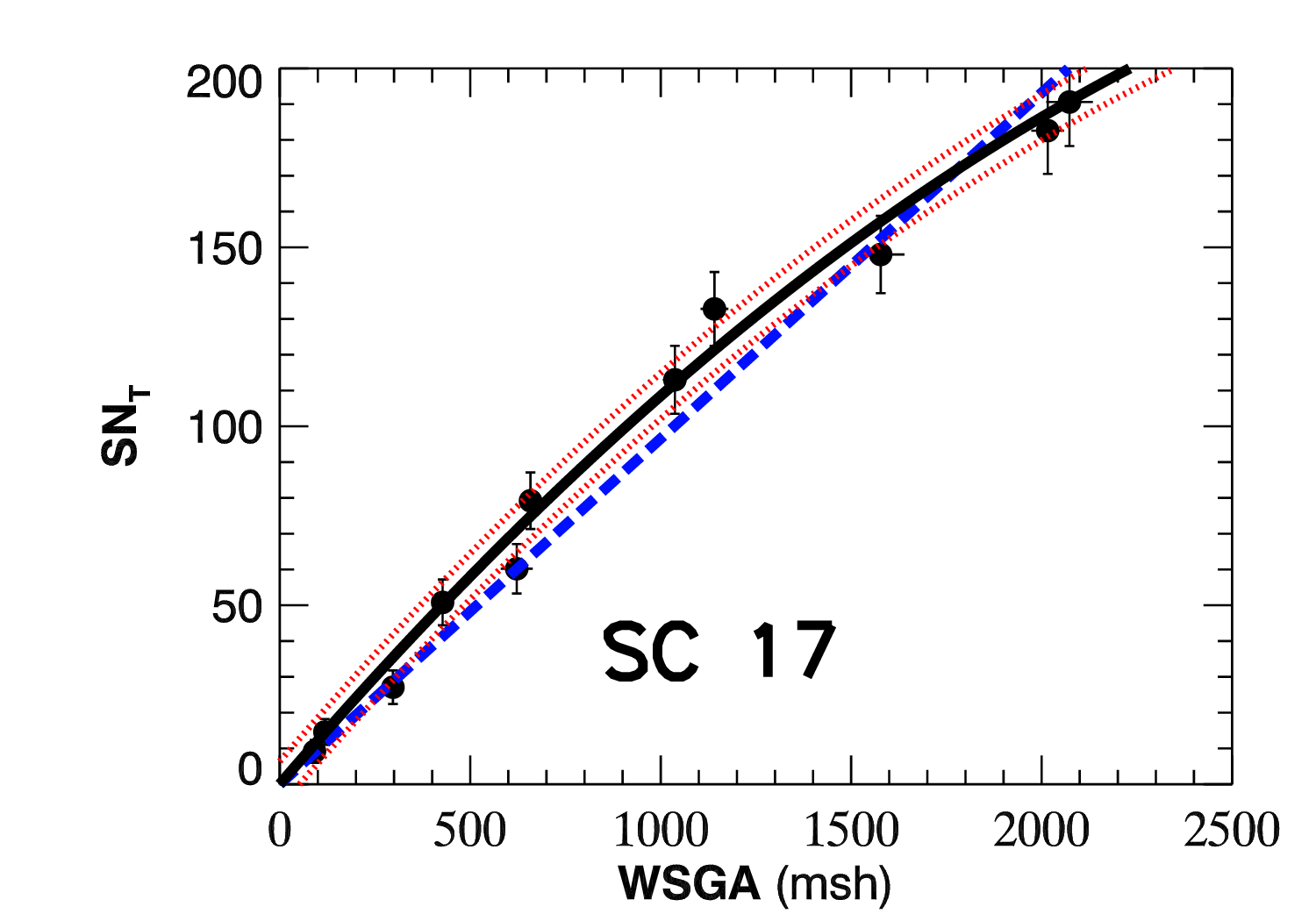}
\includegraphics[width=3.5cm]{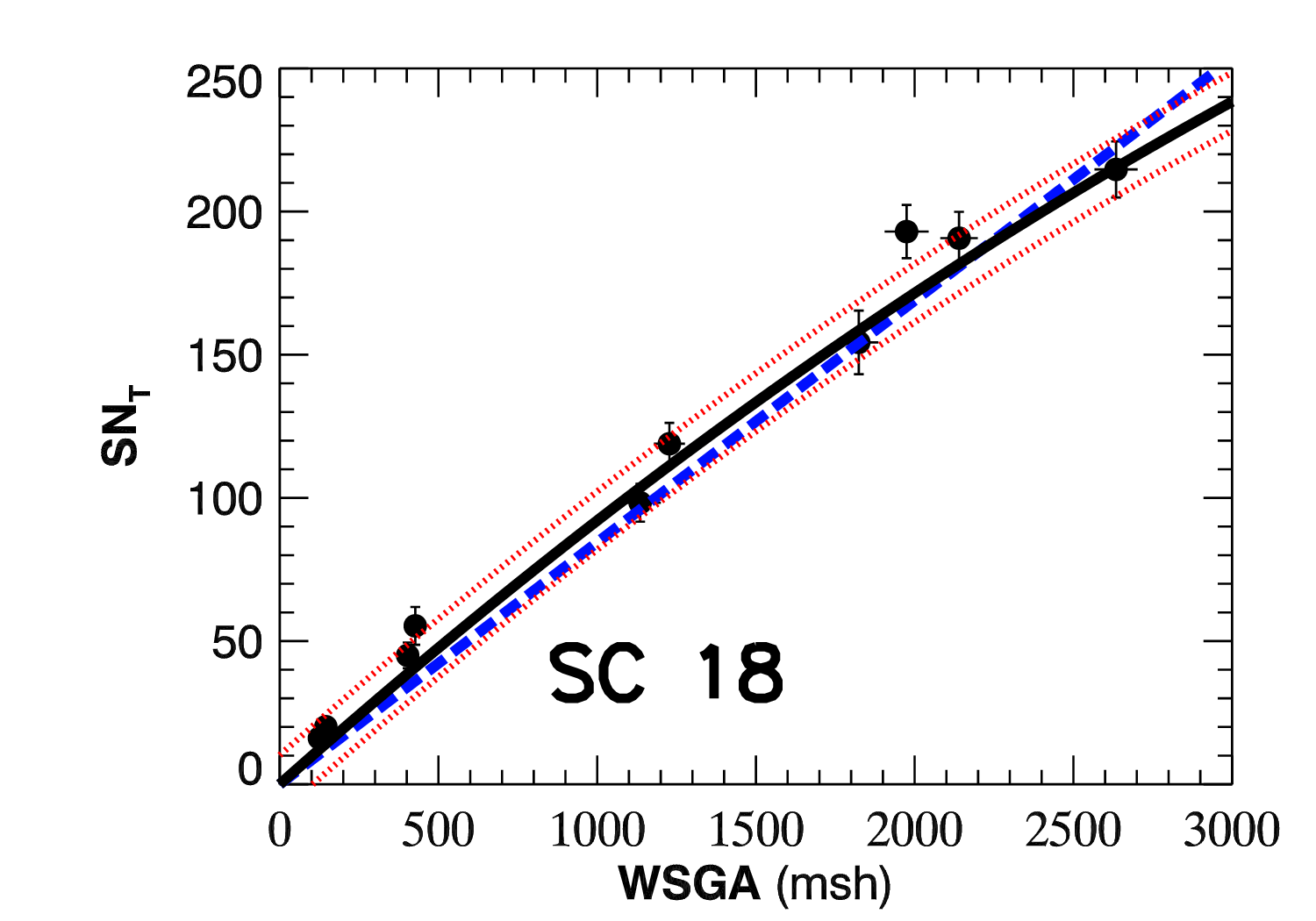}
\includegraphics[width=3.5cm]{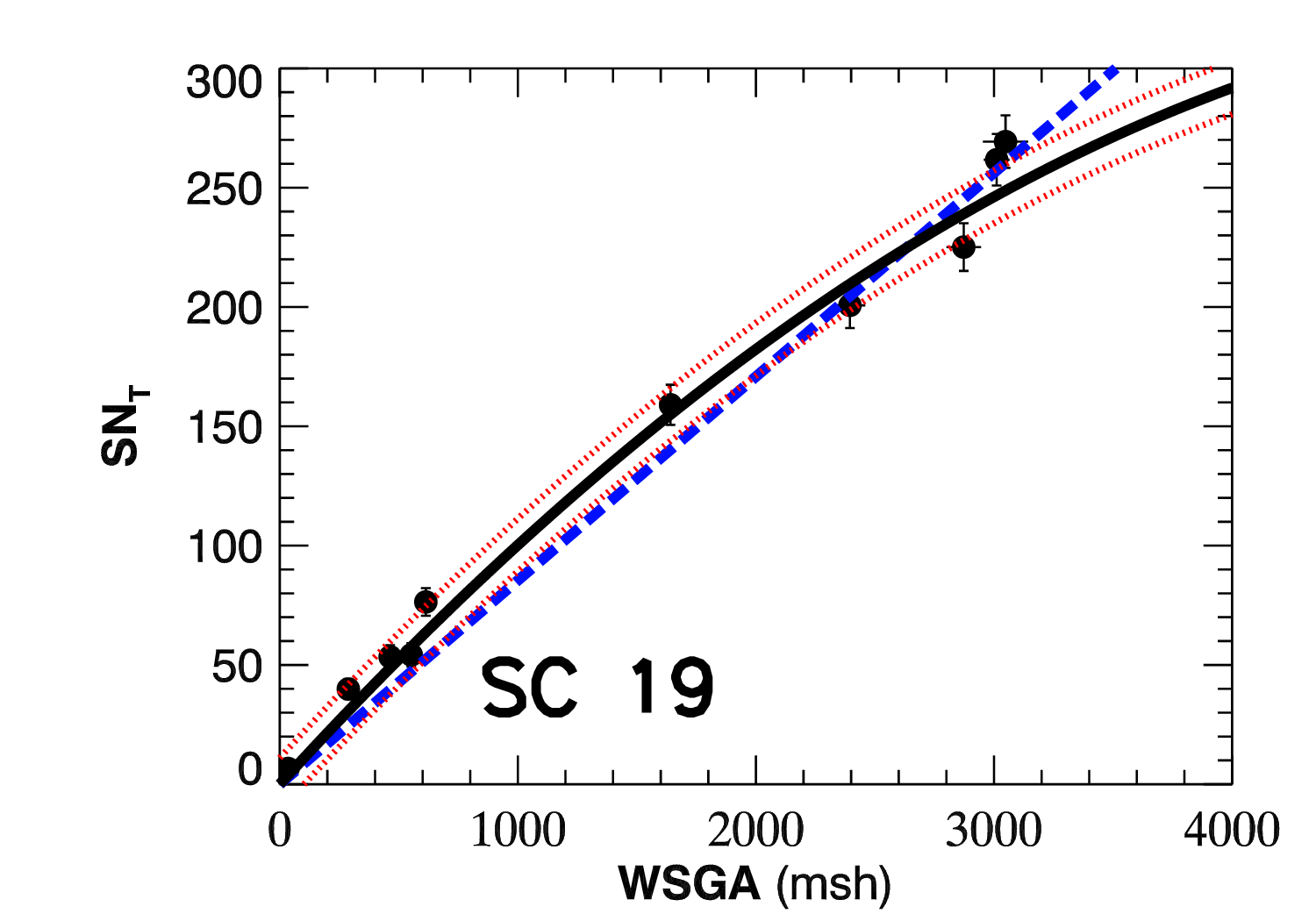}
\includegraphics[width=3.5cm]{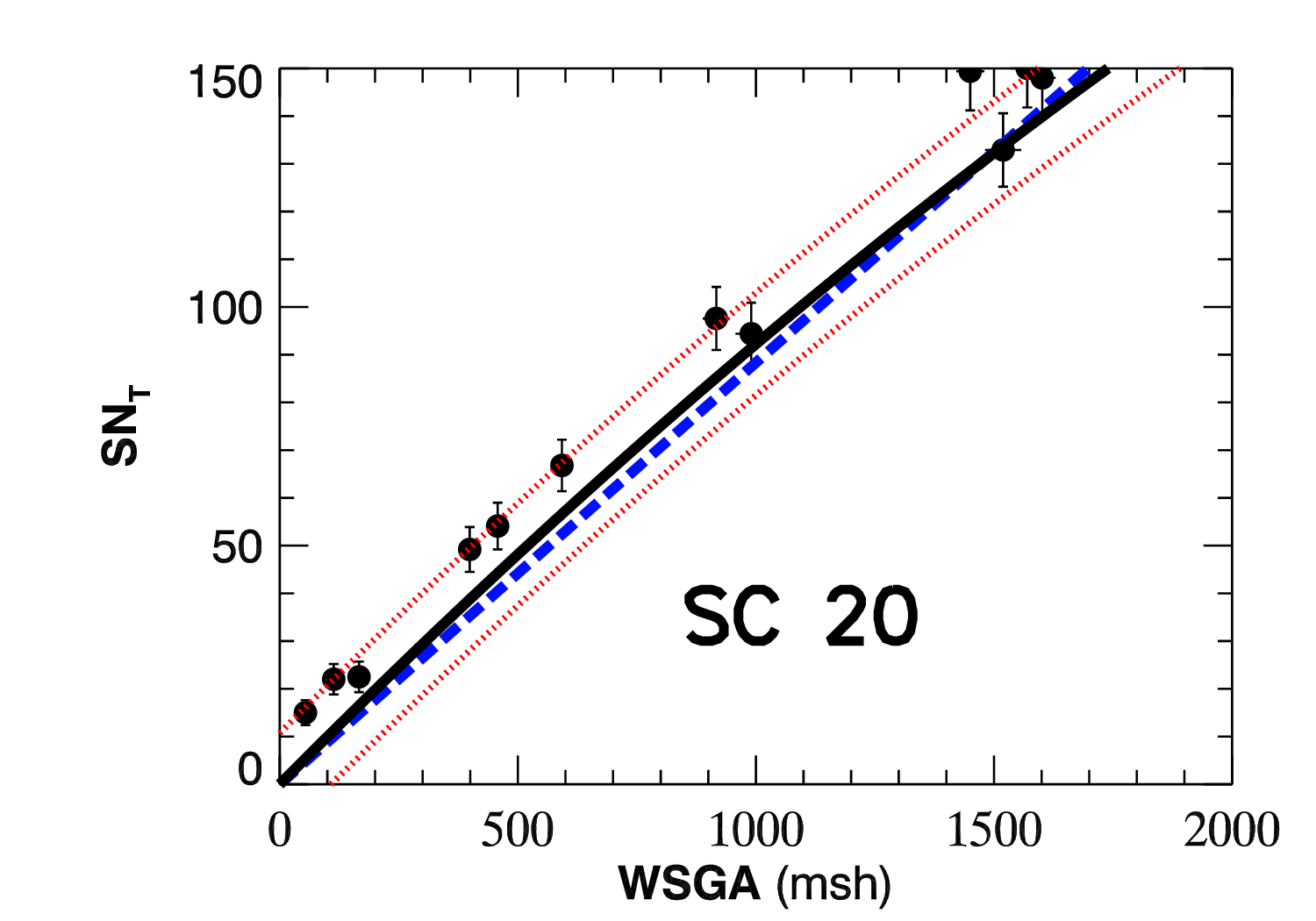}
\includegraphics[width=3.5cm]{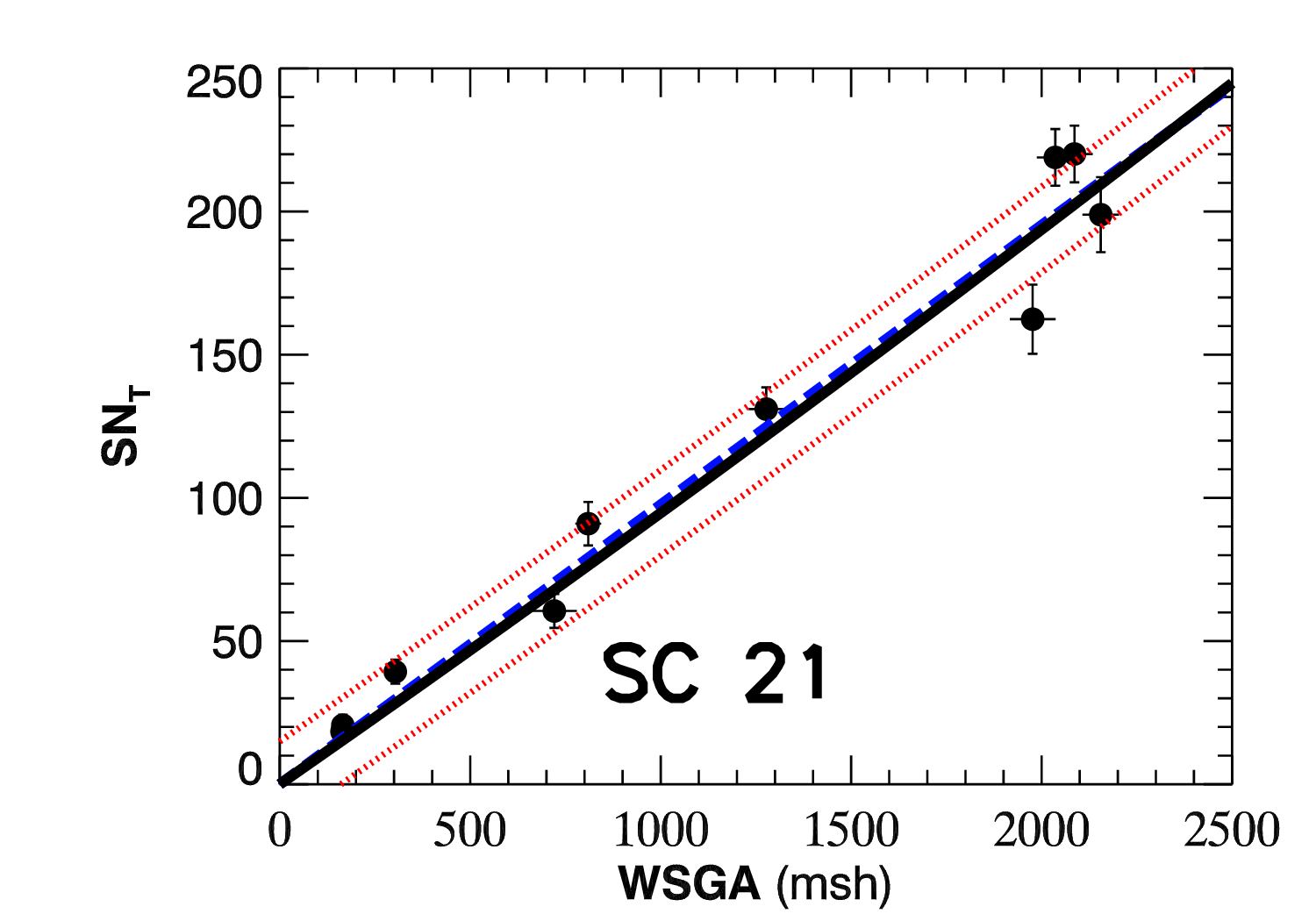}
\includegraphics[width=3.5cm]{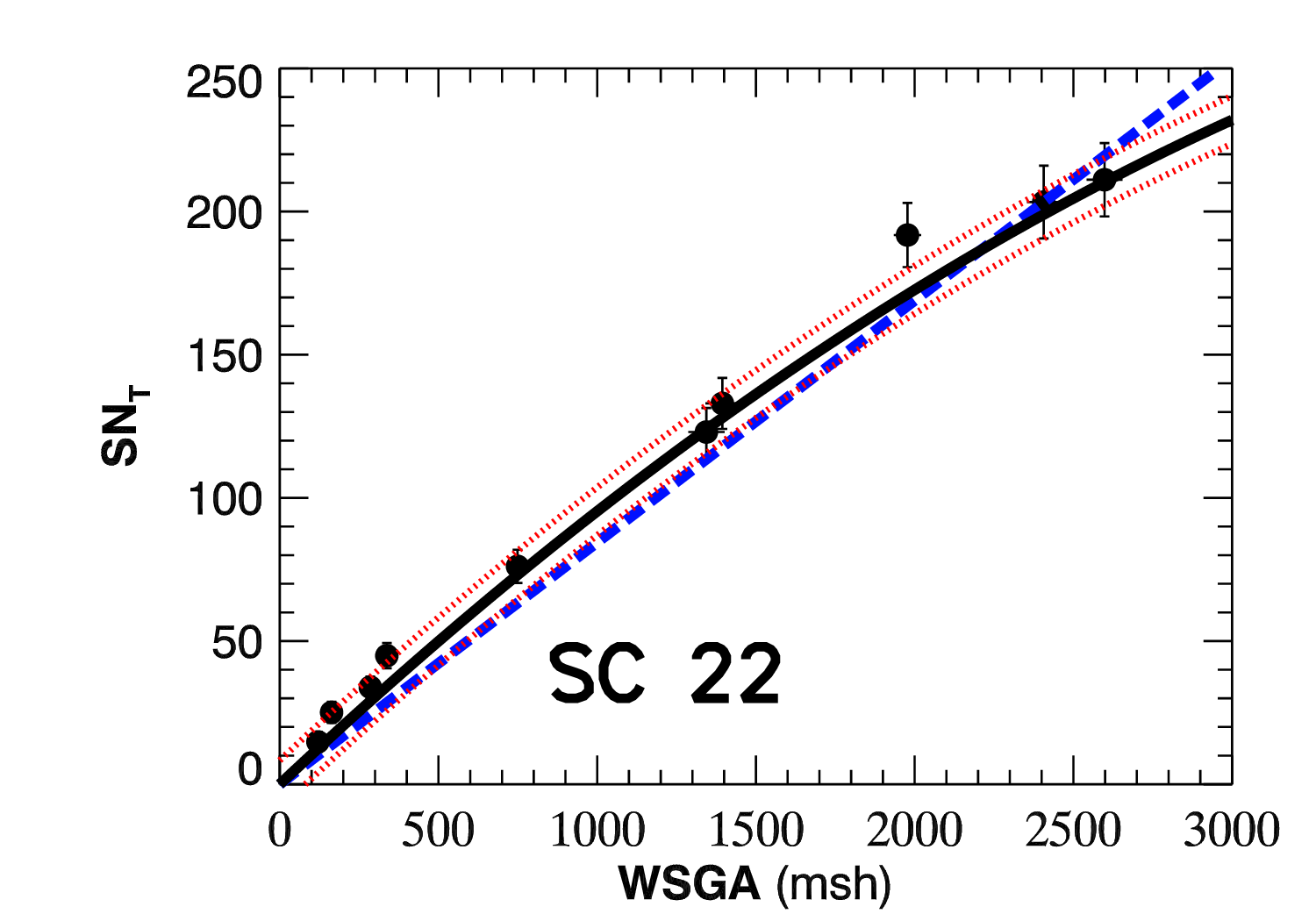}
\includegraphics[width=3.5cm]{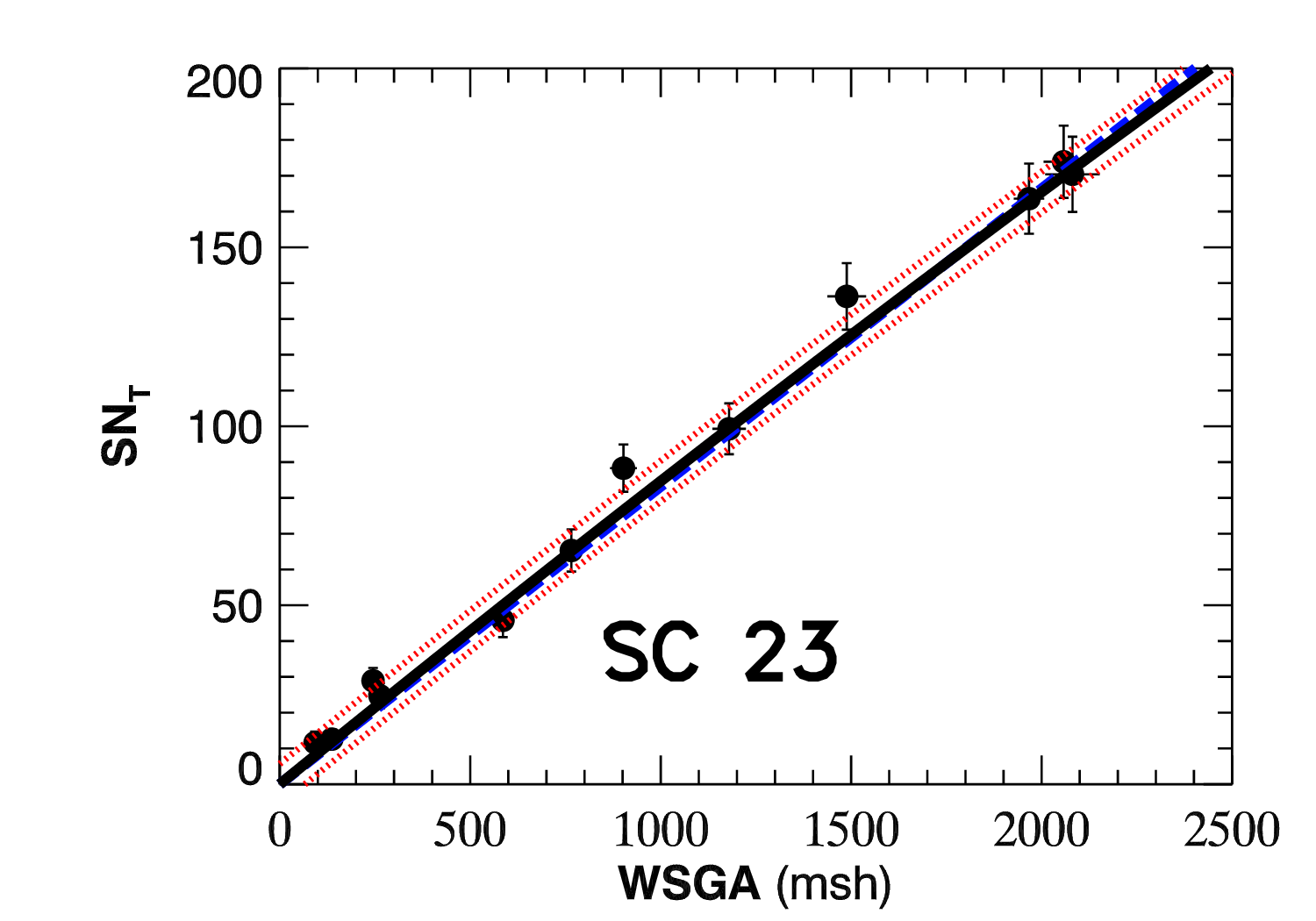}
\includegraphics[width=3.5cm]{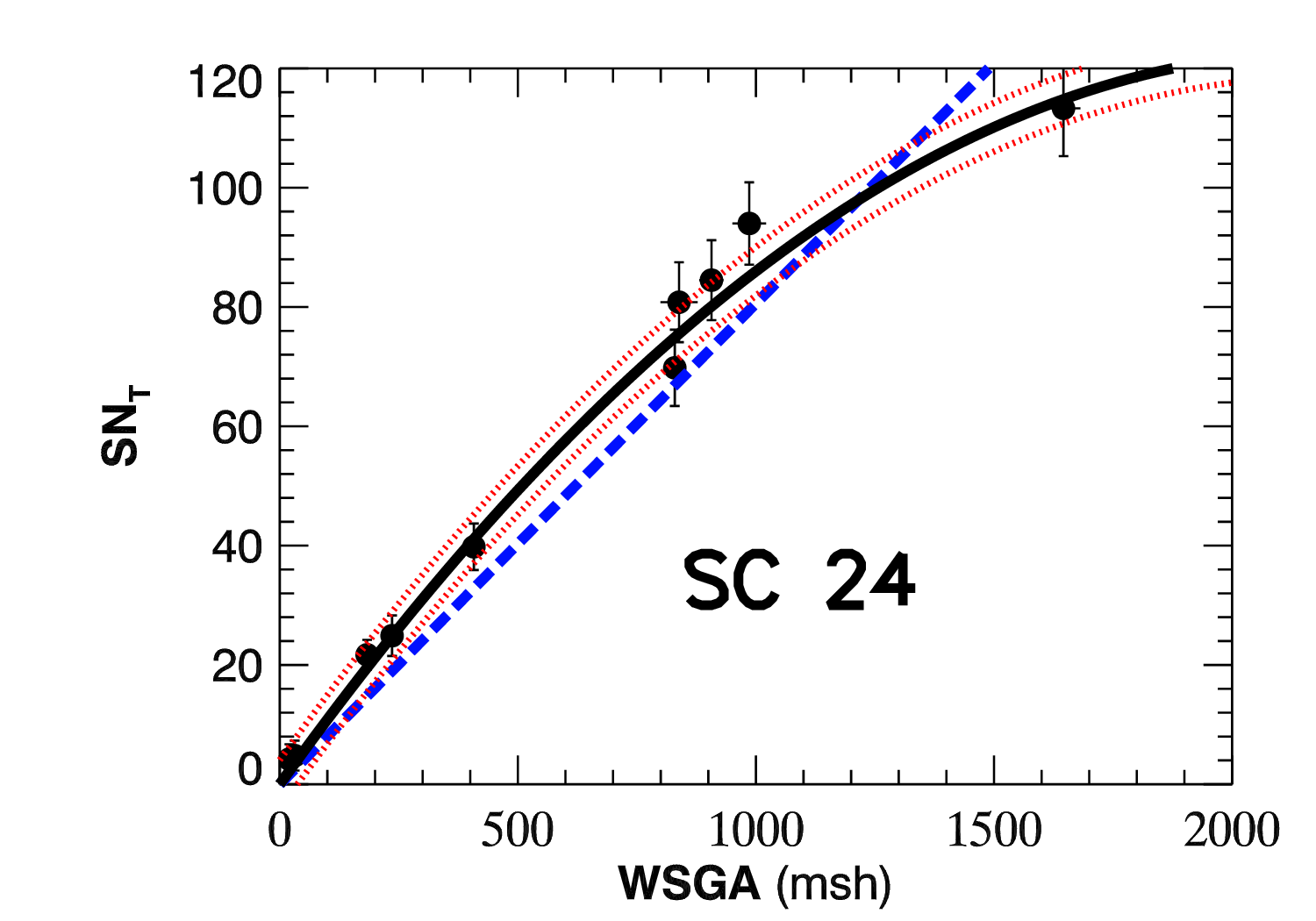}
\includegraphics[width=3.5cm]{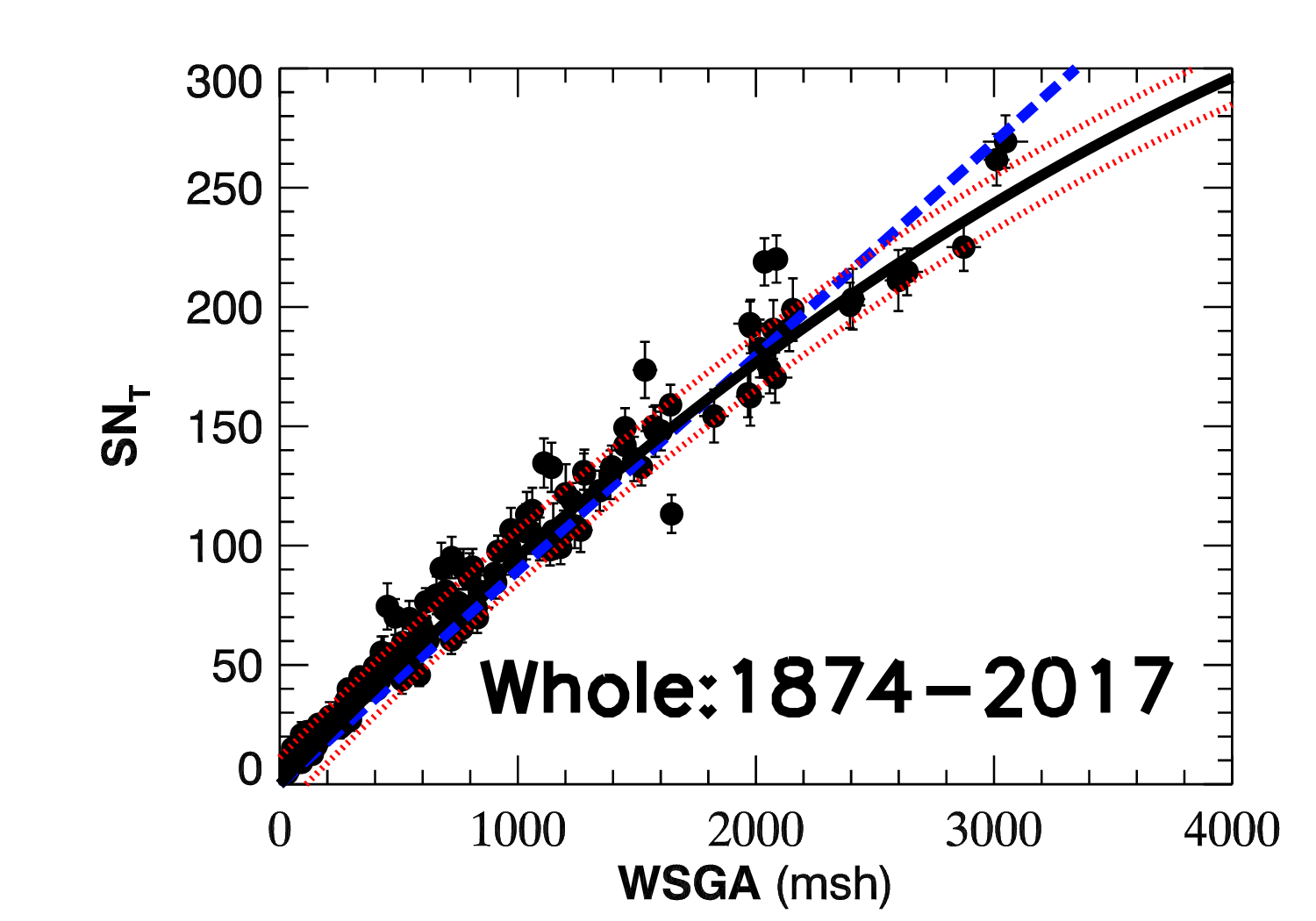}
\caption{Scatter plots  of the annual mean  WSGA versus the 
annual mean $SN_{\rm T}$ of  Solar Cycles~12\,--\,24. 
 The {\it horizontal} and {\it vertical error bars} represent the
 errors in WSGA and $SN_{\rm T}$, respectively.
The  blue {\it dashed-line} represents best-fit
 linear relationship between WSGA and $R_{\rm M}$. 
 The black {\it continuous-curve} represent the best fit
 non-linear relation and the  red {\it dotted-curve}
 represents the corresponding one-rms (root-mean-square  deviation)
 level.} 
\label{f2}
\end{figure}

 Fig.~\ref{f2} shows the  relationship between  
 $SN_{\rm T}$ and WSGA  during each of Solar Cycles~12\,--\,24 
and also during the whole period 1874\,--\,2017 
determined by using the annual mean values of 
$SN_{\rm T}$ and WSGA (in the case of Solar Cycle~24
 the data are incomplete). 
In this figure we have shown  the best fit 
linear and nonlinear relations and  the details such as 
values of slope/coefficients, correlation coefficient, $\chi^2$ and 
the corresponding probability ($P$), etc., are given in
 Tables~\ref{table1} and \ref{table2}.  Table~\ref{table1} also 
contains the values of  $R_{\rm M}$ and the mean WSGA (13-month 
smoothed monthly mean values)  at the 
maximum epoch and the corresponding uncertainties $\sigma_{\rm R}$ and
 $\sigma_{\rm A}$, respectively, of solar cycles. 
The linear least-square fit
 is calculated by taking into account  the errors in the values of 
$SN_{\rm T}$, whereas the nonlinear fit is done by taking 
 into account   the errors in the values of $SN_{\rm T}$ only. 
 Note that a small value of $P$ indicates a poor fit 
(large value of $\chi^2$). 
In the cases of all solar cycles and the whole period the correlation 
is good (significant on more than 99\,\% confidence level).
 As can be seen in Fig.~\ref{f2} and Table~\ref{table1}, 
 in all solar cycles and whole-period the 
obtained linear relations of WSGA and $SN_{\rm T}$ are good, in the 
sense that the ratio of slope to its uncertainty is substantially large.
However, in the cases of several solar cycles (14, 17, 18, 19, 21, and 24) 
and whole period the value of $\chi^2$ is large ($P$ is small), i.e. $\chi^2$ 
is significant on more than 95\,\% confidence level.   
As can be seen in Fig.~\ref{f2} and Table~\ref{table2} in the nonlinear case 
 the ratio of the coefficient ($m_1$) of the first order term to 
the corresponding uncertainty  is reasonably large in  all solar cycles  
and the corresponding ratio of the second order term is significant 
 only in a few cycles (14, 17, and 24). In these cycles the $\chi^2$ is 
also insignificant. That is, for each of these cases  
the nonlinear fit is better than that of the linear fit. 
 In the case of the whole period
 the $\chi^2$ value of the linear fit is slightly smaller than that of the 
  nonlinear fit, i.e. in this case the linear fit seems to be slightly better
 than nonlinear fit. 
(Similar results are also found from the monthly mean and the 13-month 
smoothed monthly mean values of WSGA and $SN_{\rm T}$).  

\begin{figure}
\centering
\includegraphics[width=8.5cm]{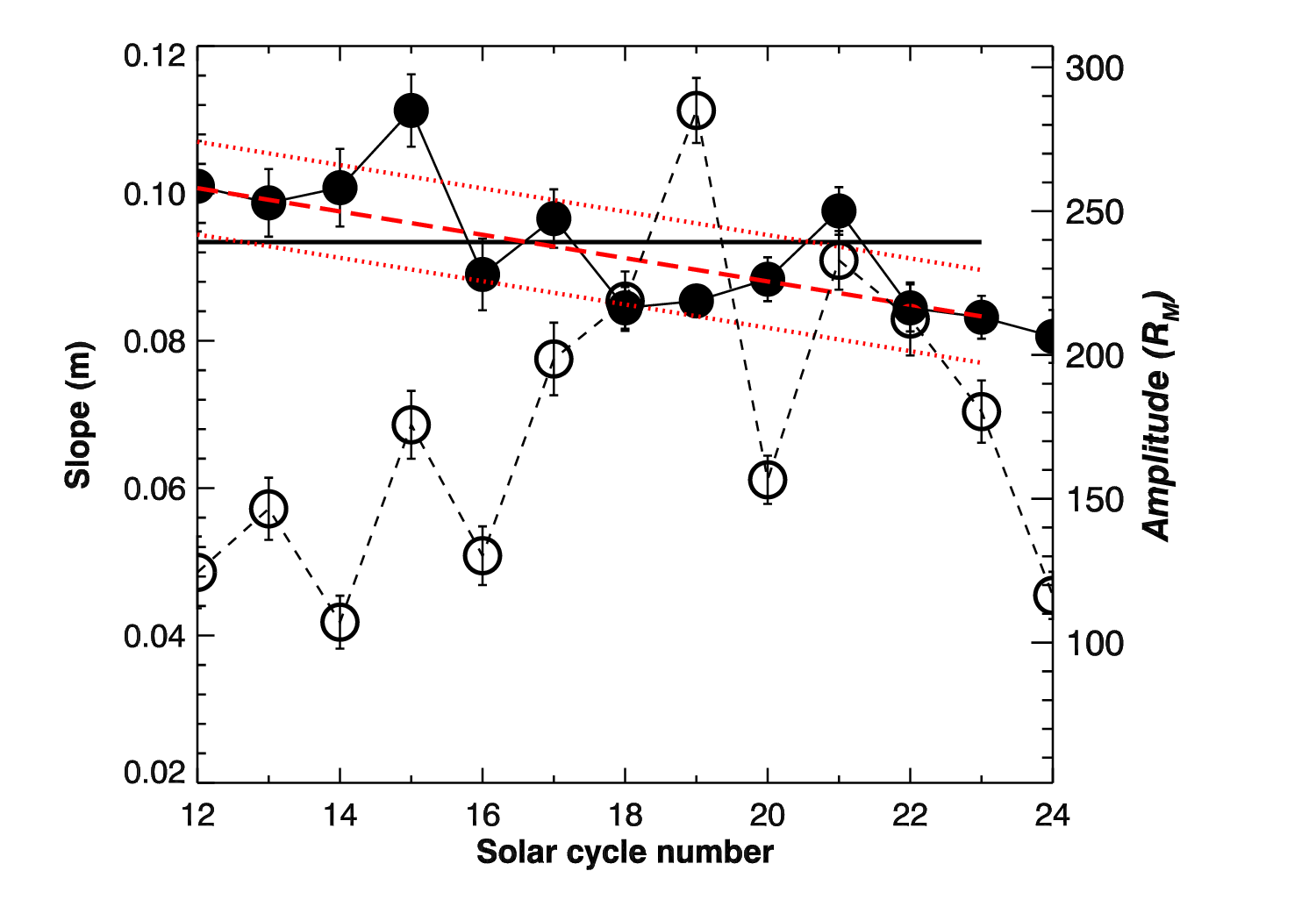}
\caption{The {\it filled circle-continuous curve} represents solar
 cycle-to-cycle variation in the slope ($m$) of the linear
relationship (Eq.~(\ref{eq1})) of WSGA and $SN_{\rm T}$ during 
 Solar Cycles~12\,--\,24, 
 determined by using the corresponding annual mean values.
 The {\it horizontal continuous-line}  represents the mean value of the slopes
 during Solar Cycles~12\,--\,23. The {\it long-dashed line} (red) represents
 the secular decreasing  trend in the slope obtained from the best-fit linear
 relation (Eq.~(\ref{eq3})) between the slope and  solar cycle number during
 Solar Cycles~12\,--\,23, and the corresponding one-rms level is shown by the
 red {\it dotted-line}. The open circle-dashed curve represents the
variation in the amplitude ($R_{\rm M}$) of solar cycle. The data point of the
incomplete Solar Cycle~24 is not used for the determination of both the mean
and the linear-least-square fit.}
\label{f3}
\end{figure}

Fig.~\ref{f3} shows the cycle-to-cycle variation in the slope ($m$)
 of the nonlinear
relation between WSGA and $SN_{\rm T}$ during Solar Cycles 12\,--\,24.
In this figure we have also shown the variation in the amplitude ($R_{\rm M}$)
of solar cycle. The values of $R_{\rm M}$
of  Solar Cycles~12\,--\,24 are
taken from \cite{pesnell18}.  As we can see in this figure there exists a
considerable variation in the slope. The slope of the average Solar Cycle~15
 is largest and is at about $2\sigma$ (standard deviation) level (significant
 on about 95\% confidence level). No significant correlation is found between
 the slope and the amplitude ($R_{\rm M}$) of a solar cycle. There is a
 suggestion of the existence of a secular decreasing trend in the slope from
Solar Cycle~12\,--\,24 and it seems superimposed on it a weak 77-year long-term 
 variation (may be related to the Gleissberg cycle).
 We obtained the following linear relationship between the slope ($m$)
 and  solar cycle number ($n$).
\begin{equation}
m = 0.12\pm0.0065 - (0.00158\pm0.00034)n
\label{eq3}
\end{equation}
The corresponding  correlation  of this relation (Eq.~(\ref{eq3})) is not very
good ($r = -0.69$), but the least-square fit is reasonably good, i.e. the slope
 is reasonably well defined (the ratio of the slope to its
standard deviation is about 5) and the rms (root-mean-square  deviation)
 value 0.0063 is reasonably small
(the data point of the incomplete Solar Cycle~24 is excluded for the
 determination of both the mean and the linear-least-square fit.)

\begin{figure}
\centering
\includegraphics[width=8.5cm]{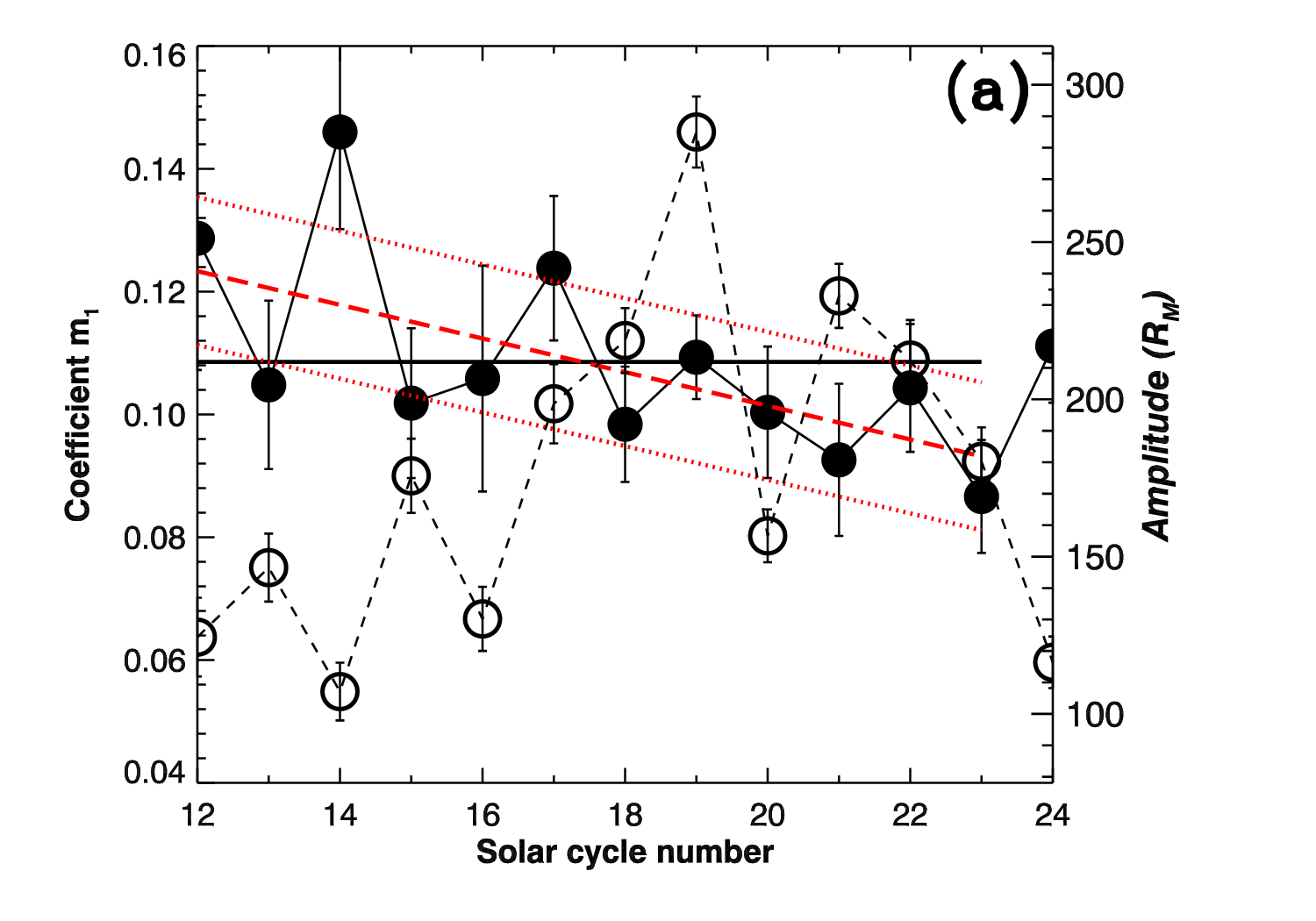}
\includegraphics[width=8.5cm]{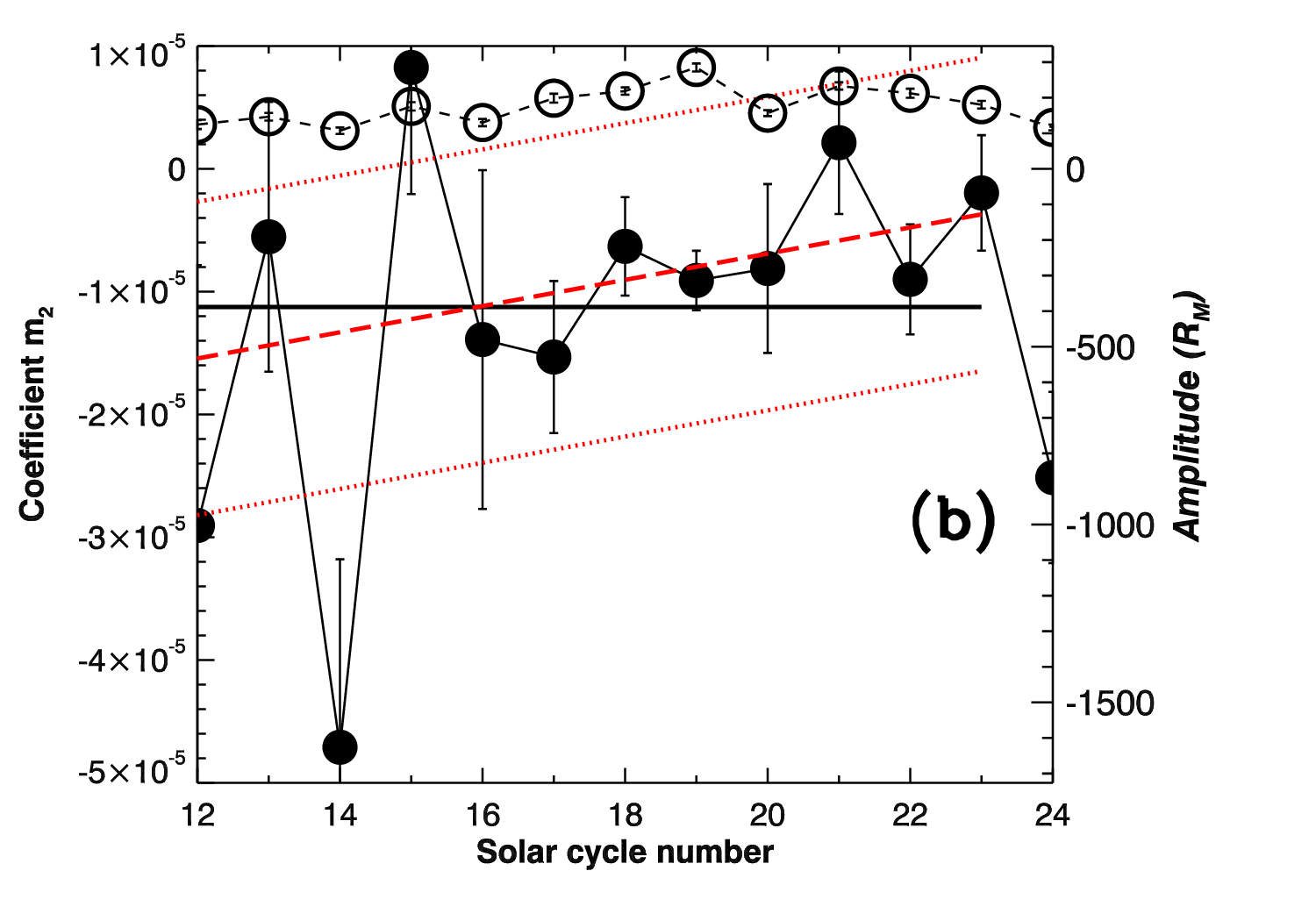}
\caption{The {\it filled circle-continuous curve} represents solar
 cycle-to-cycle variation in  ({\bf a})  the coefficient $m_1$ and 
({\bf b})  the coefficient $m_2$ of the nonlinear 
relationship (Eq.~(\ref{eq2})) of WSGA and $SN_{\rm T}$
 during the whole Solar Cycles~12\,--\,24,
 determined by using the corresponding annual mean values.
 The horizontal continuous line represents the mean value of the coefficients 
 during Solar Cycles~12\,--\,23. The long-dashed line (red) represents the
 secular trend in the coefficient obtained from the best-fit linear
 relation (Eq.~(\ref{eq4})) between $m_1$ 
 and the  solar cycle number ($n$), and between $m_2$ and $n$ (Eq.~(\ref{eq5}), 
during  Solar Cycles~12\,--\,23, and the corresponding one-rms level is
 shown by the red dotted-line. The open circle-dashed curve represents the 
variation in the amplitude ($R_{\rm M}$) of solar cycle. The data point of the 
incomplete Solar Cycle~24 is not used for the determination of both the mean 
and the linear least-square fit.}
\label{f4}
\end{figure}

Figs.~\ref{f4}a and \ref{f4}b show
 the cycle-to-cycle variations in the coefficients 
 $m_1$ and $m_2$ of the    
 nonlinear relation between WSGA and $SN_{\rm T}$ during Solar Cycles 12\,--\,24.
In these figures also we have  shown the variation in the
 amplitude ($R_{\rm M}$)
of solar cycle. As we can see in this figure there exists  
considerable variations in both the coefficients $m_1$ and $m_2$. 
The $m_1$ of Solar Cycle~14 
 is largest and is at about $2\sigma$ (standard deviation) level, 
i.e. significant on about 95\% confidence level  
(note that in the case of the linear relation 
 $m$ of Solar Cycle 15 is largest).   
 No significant correlation is found between
 the any of these coefficients and the amplitude ($R_{\rm M}$) of a solar cycle.
As in the case of the slope $m$ of Eq.~(\ref{eq1}) shown in Fig.~\ref{f3},
   there is a suggestion of the existence of a secular decreasing trend in
  $m_1$, but there seems to be a  secular increasing trend in $m_2$
 from Solar Cycle~12\,--\,24.
 We obtained the following linear relations between $m_1$ and 
  solar cycle number ($n$) and between $m_2$ and $n$.
\begin{equation}
m_1 =0.16 \pm0.02 - (0.0027\pm0.0011)n,
\label{eq4}
\end{equation}
\begin{equation}
m_2 = -2.82\times10^{-5}\pm 1.33\times10^{-5} + (1.07\times10^{-6}\pm 6.85\times10^{-7})n.
\label{eq5}
\end{equation}
The corresponding  correlation  of Eq.~(\ref{eq4}) is not very 
good ($r = -0.65$), but the least-square fit is reasonably good, i.e. the
ratio of the  slope of Eq.~(\ref{eq4})  to its 
standard deviation is about 2.5 and the rms value 0.012 is reasonably small
(the data point of the incomplete Solar Cycle~24 is excluded for the
 determination of both the mean and the linear-least-square fit.)
The corresponding correlation of Eq.~(\ref{eq5}) is small ($r = 0.48$), the 
ratio of $m_2$ its standard deviation is only 1.47, and rms value is
 $1.28\times10^{-5}$. Overall, this relation seems to be not good.

\begin{figure}
\centering
\includegraphics[width=9cm]{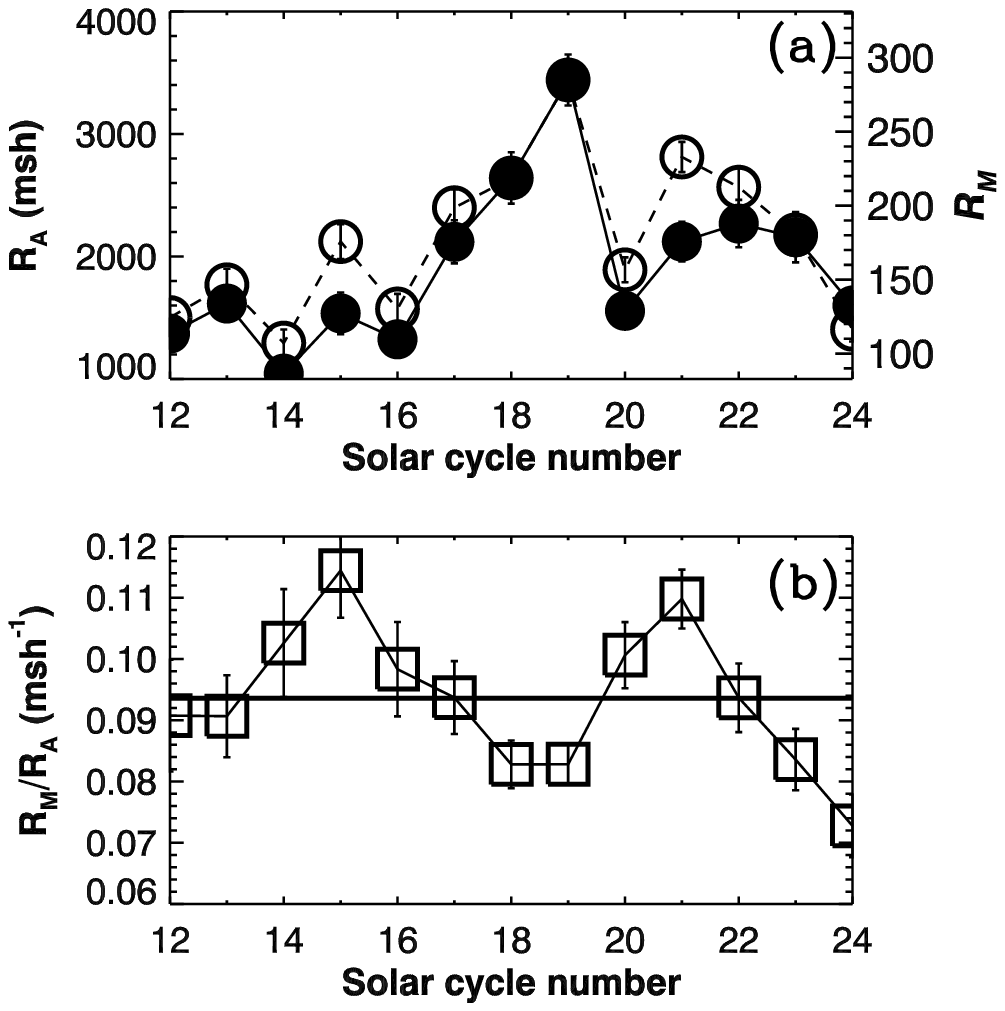}
\caption{({\bf a}) The {\it filled circle-continuous curve} represents 
 variation in $R_A$, i.e. the value of  the 13-month smoothed monthly mean 
area of the sunspot groups (WSGA) in whole-sphere at the 
 maximum epoch of  solar cycle and the 
 {\it open circle-dashed curve} represents
 the amplitude ($R_{\rm M}$) of  solar cycle, during Solar Cycles 12\,--\,24.
 ({\bf b}) Solar cycle-to-cycle variation of 
the ratio $R_{\rm M}/R_A$ ({\it square-continuous curve}).
 The {\it horizontal line} represents the mean value (0.0936) of 
 $R_{\rm M}/R_A$ over  Solar Cycles 12\,--\,24.} 
\label{f5}
\end{figure}

In Fig.~\ref{f5}a we have shown the variations in the values of  $R_A$  
 (the value of the 13-month smoothed monthly mean WSGA at the maximum epoch
 of a solar cycle) and $R_{\rm M}$ during  Solar Cycles 12\,--\,24 
(values are given in Table~1).
The patterns of variations of $R_{\rm M}$  and $R_A$ are almost the same, 
except in the case of the former the value of Solar Cycle 21 is larger 
than that of Solar Cycle~22, but it is opposite in the case of the latter.
 Fig.~\ref{f5}b shows the solar cycle-to-cycle  variation of the ratio 
$R_{\rm M}/R_A$. The variation of the ratio  is very closely  
resembles to that of the slope $m$ (see Fig.~\ref{f3}). There 
exists a reasonably good correlation ($r = 0.7$) between the slope and the
 ratio.  A $\approx$77-year  variation of the ratio
 can be seen better than that of the slope $m$  (similar result 
is also found from the annual mean values of $R_A$ and $R_{\rm M}$). 
The mean $R_{\rm M}/R_A$ is 0.0936 and the corresponding 
standard deviation ($\sigma$)  
 is 0.0117. The amplitude of the 77-year variation is $\approx$1.79$\sigma$.
 That is, the amplitude  is only slightly less than 
95\,\% confidence level. Based on the locations, 
  at Solar Cycles $n = 15$ and $n = 21$, of the crests of this  
 variation  we can make the following predictions.
 
Let us consider the  crest at  either Solar Cycle $n =15$ or 
at Solar Cycle $n =21$. 
  We can see that:\\
 (i) Solar Cycle $n-1$ (14/20) is  smaller 
(in amplitude) than  Solar Cycle $n$ (15/21),\\
 (ii) Solar Cycle $n-2$ (13/19) is  larger than  Solar Cycle $n-1$ (14/20),\\
 (iii) Solar Cycle $n-3$ (12/18) is  smaller than  Solar Cycle $n-2$ (13/19),
 and \\ 
(iv) the behavior of the ratio $R_{\rm M}/R_A$ at Solar Cycle 27 is expected 
 to be similar as those of Solar Cycles 15 and 21.

Using the aforementioned pattern we can expect Solar Cycle~26 will be  
 smaller than Solar Cycle~27 (from (i) and (iv) above),
 Solar Cycle 25 is  larger than Solar Cycle~26 (from (ii) and (iv) above),  
Solar Cycle 24 is  smaller than Solar Cycle~25 (from (iii) and (iv) above), 
i.e. Solar Cycle 25 will be  either very large as Solar Cycle~19  or
 it is below a average solar cycle as Solar Cycle~13 but larger than Solar 
Cycle 24, and  the solar cycle pair (26, 27) will satisfy the Gnevyshev-Ohl
 rule or G-O rule  of solar cycles \citep{go48}.

Based on  the pattern of the variation in $R_{\rm M}/R_A$ we can also notice 
that Solar Cycle $n+1$ (16/22) is smaller than Solar Cycle $n$ (15/21)
 suggesting that Solar Cycle 28 will be smaller than Solar Cycle 27.
However, here the relative strengths of Solar Cycles $n+1$ and $n+2$ can not be 
predicted  because the solar cycle pair (16, 17) satisfied the G-O rule, 
whereas the solar cycle pair (22, 23) violated the G-O rule. Hence, the 
relative size of Solar Cycle 29 with respect to that of Solar Cycle~28 is
 not possible to predict from the pattern of the variation in 
the ratio $R_{\rm M}/R_A$.          

\subsection{An improved prediction for the amplitude of solar cycle 25}
 Here we refer to REL-I and REL-II (for definitions see Sec. 2).
 In most of our earlier papers \citep{jj08,jj15,jj17,jj21,jj23} we have
 mentioned that $T^*_{\rm M}$ of REL-I of a solar cycle  is
related to the Gnevyshev gap of that cycle.
It should be noted that as can be seen Fig.~1
 in a majority  of solar cycles among the Gnevyshev peaks, the  
major peak is occurred first and the secondary peak is occurred later (second).
 In a few solar cycles this is opposite. However, only in a few solar cycles
  Gnevyshev peaks and  Gnevyshev gapes are well defined. In Solar Cycles 
23 and 24 the Gnevyshev gapes are well defined, i.e. well separated but 
the secondary peak is occurred first and the major peak is occurred later.
Such instances cause errors in the determination of $A^*_{\rm M}$ and 
increases inconsistency of REL-I. This is because according to the
 definition of REL-I the epoch $T^*_{\rm M}$ of a solar cycle is
 just after the maximum epoch.  In our earlier analyses in the cases of 
Solar Cycles 23 and 24 also we  have used the  $T^*_{\rm M}$ after the
 major peaks (second peaks). This might  be  a reason for earlier we obtained
 a substantial low value for $A^*_{\rm M}$ of Solar Cycle 24 and a
 low/incorrect value for the amplitude of Solar Cycle~25.
 Here we analysed the GPR and DPD sunspot group data during 
1874\,--\,2017 and obtained a revised REL-I by 
considering the epochs of the first peaks (instead of the epochs of 
 the major peaks) of Solar Cycles 23 and 24. We have also 
determined REL-II (note that no aforementioned problems 
in the case of REL-II). 

\begin{figure}
\centering
\includegraphics[width=8.0cm]{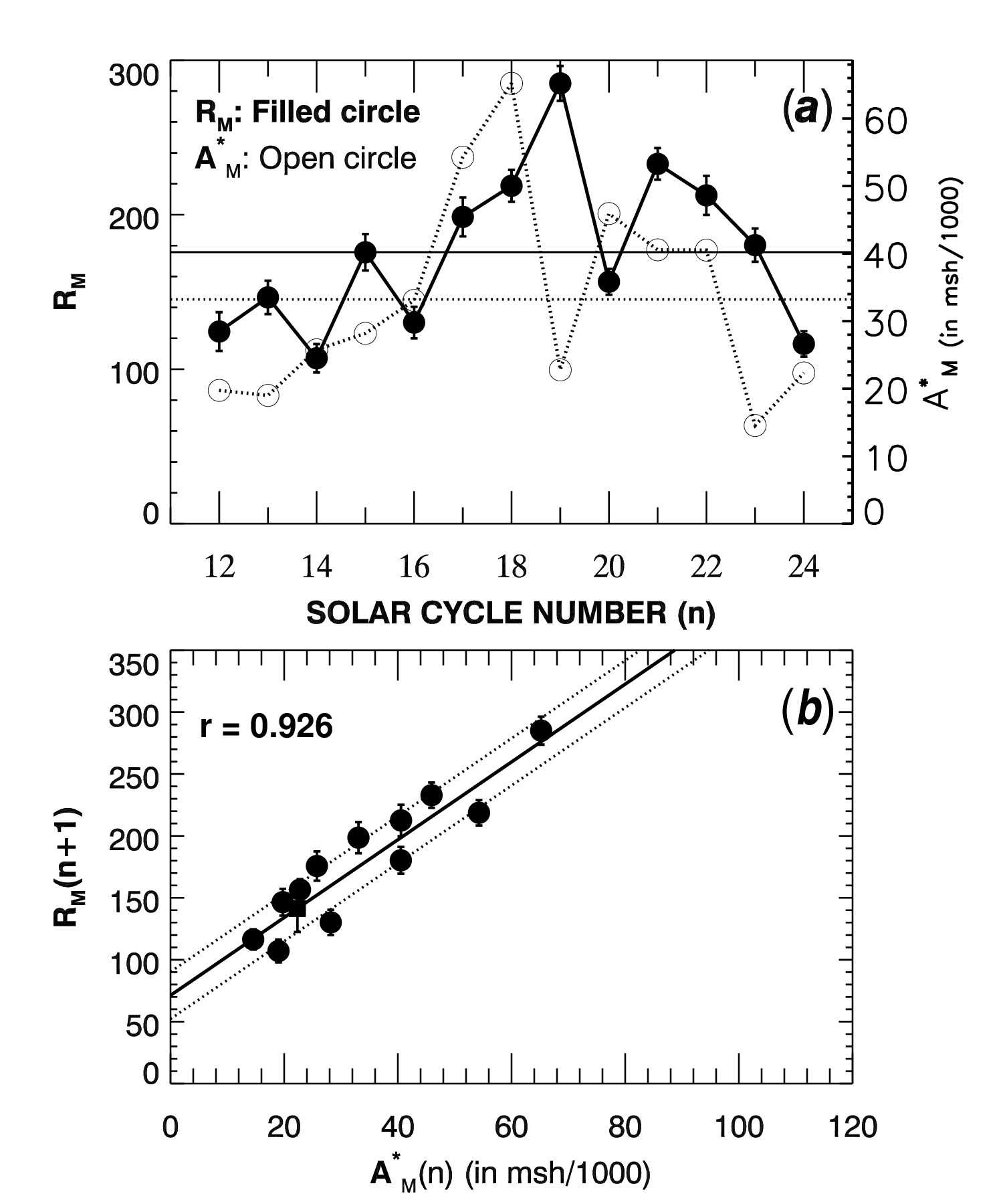}
\caption{({\bf a}) Plot of the amplitude  ($R_{\rm M}$,
i.e. the largest 13-month smoothed monthly mean $SN_{\rm T}$)  of
 a solar cycle and the sum of the areas ($A^*_{\rm M}$) of the sunspot groups
in $0$\,--\,$10^\circ$ latitude interval of the Sun's southern hemisphere
during the interval ($T^*_{\rm M}$) around the maximum  epoch ($T_{\rm M}$)
of the solar cycle versus the solar cycle number ($n$).
({\bf b}) The scatter plot of $A^*_{\rm M}$ of a Solar Cycle $n$  and 
 $R_{\rm M}$  of Solar Cycle $n+1$. The {\it continuous line}
represents  the corresponding best-fit linear relation REL-I
  (Eq.~(\ref{eq6})) and the {\it dotted lines}  are drawn at one-rms level. 
 The values of the correlation coefficient ($r$) is also shown.
 The {\it filled-square} represents
the predicted value $141 \pm 19$ of $R_{\rm M}$ of Solar Cycle~25.}
\label{f6}
\end{figure}

\begin{figure}
\centering
\includegraphics[width=8.0cm]{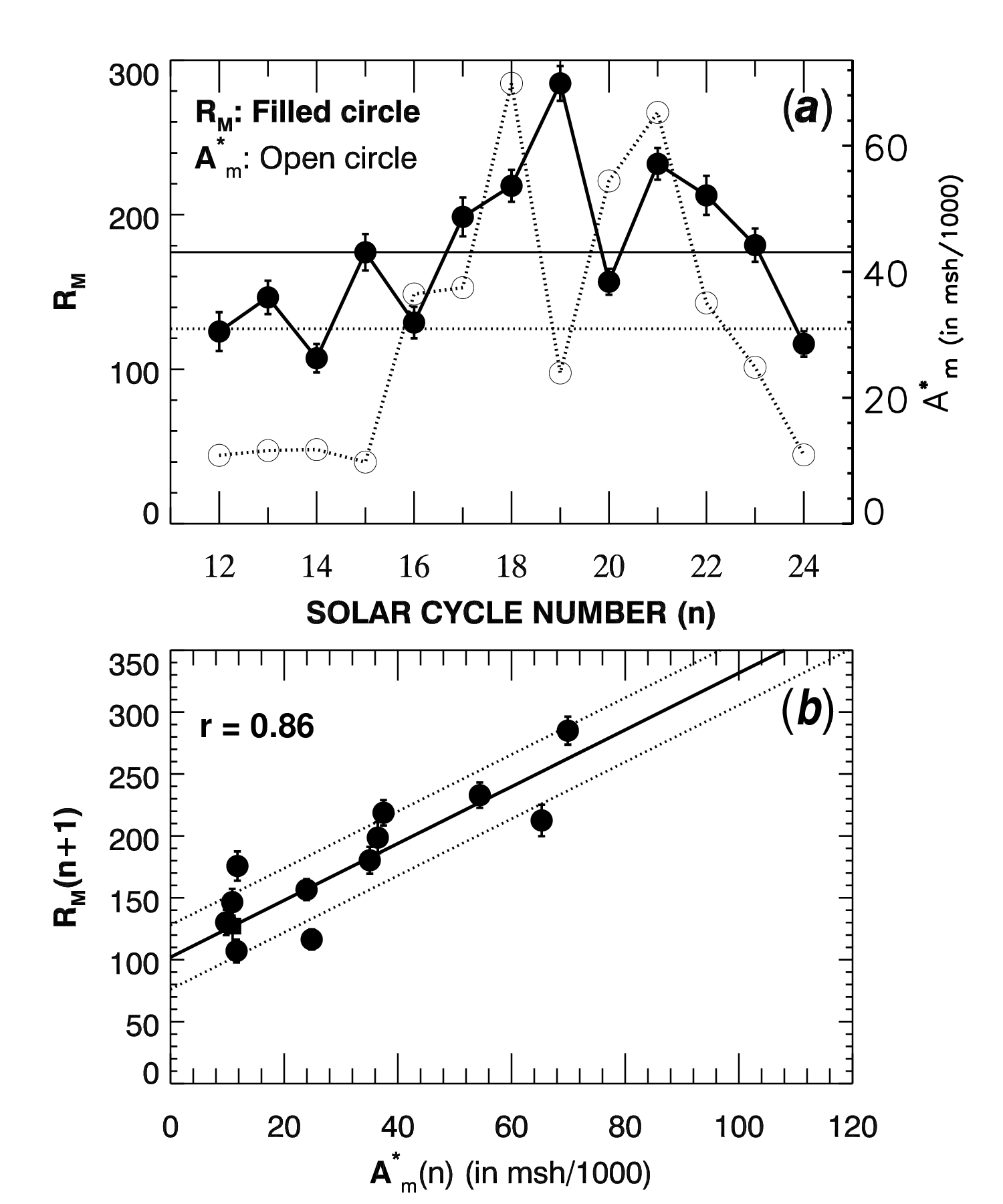}
\caption{({\bf a}) Plot of the amplitude  ($R_{\rm M}$)
 of a solar cycle and the sum of the areas ($A^*_{\rm m}$) of 
the sunspot groups
in $0$\,--\,$10^\circ$ latitude interval of the Sun's northern hemisphere
during the interval ($T^*_{\rm m}$) around the minimum  epoch ($T_{\rm m}$)
of the solar cycle versus the solar cycle number ($n$).
({\bf b}) The scatter plot of $A^*{\rm_m}$ of a Solar Cycle $n$  and 
 $R_{\rm M}$  of Solar Cycle $n+1$. The {\it continuous line}
represents  the corresponding best-fit linear relation REL-II
 (Eq.~(\ref{eq7})) and the {\it dotted lines}  are drawn at one-rms
  level.  The values of the correlation
 coefficient ($r$) is also shown.
 The {\it filled-square} represents
the predicted value $127 \pm 26$ of $R_{\rm M}$ of Solar Cycle~25.}
\label{f7}
\end{figure}

 In Table~\ref{table3}  we have given the values 
 obtained here for the intervals $T^*_{\rm _M}$ and $T^*_{\rm _m}$  
of the Solar Cycles 12\,--\,24 and the values of the sums of the 
areas, $A^*_{\rm M}$ and $A^*_{\rm m}$,  of the sunspot groups during
 these intervals in the  $0$\,--\,$10^\circ$ latitude intervals of 
the southern and northern  hemispheres, respectively.
 Fig.~\ref{f6}a
shows the cycle-to-cycle variations in  $A^*_{\rm M}$
  and  $R_{\rm M}$ during  Solar Cycles 12\,--\,14.  
Fig.~\ref{f6}b shows the relationship between $A^*_{\rm M}$ of
 a solar cycle $n$ and 
$R_{\rm M}$ of solar cycle $n+1$. We  obtained the following 
relation (revised REL-I, i.e. in the cases of Solar Cycles 23 and 24  
the epochs of the first peaks are considered).
\begin{equation}
R_{\rm M} (n+1) = (3.14 \pm 0.20) A^*_{\rm M} (n)  + 71 \pm 7.
\label{eq6}
\end{equation}
Fig.~\ref{f7}a
shows the cycle-to-cycle variations in  $A^*_{\rm m}$
  and  $R_{\rm M}$ during  Solar Cycles 12\,--\,24.  
Fig.~\ref{f7}b shows the relationship between $A^*_{\rm m}$
 of a solar cycle $n$ and 
$R_{\rm M}$ of solar cycle $n+1$. We  obtained the following 
relation  (REL-II).
\begin{equation}
R_{\rm M}(n+1) = (2.29 \pm 0.16) A^*_{\rm m} (n)  + 102 \pm 6.
\label{eq7}
\end{equation}

Both  REL-I and REL-II are derived  by taking into account 
the uncertainties in the values of the amplitudes of Solar Cycles 12\,--\,24 
\cite[for details see][]{jj07,jj21}.  
The least-square-fits of these relations are reasonably good. That is, 
the corresponding correlation ($r = 0.926$) of REL-I is statistically  
significant on more than 99.9\,\% confidence level (Student's $t=7.7$)
 and the slope of Eq.~(\ref{eq6}) is about 16 times larger than the
 corresponding standard deviation.  
By using this relation  we obtained $141\pm19$ for 
the amplitude of Solar Cycle~25. That is, we obtained  a much larger
 value than that was obtained in our earlier analyses. It is also considerably 
larger than the amplitude of Solar Cycle 24.

The corresponding correlation ($r = 0.86$) of REL-II is also statistically  
significant on 99.9\,\% confidence level (Student's $t= 5.36$) and the 
slope of Eq.~(\ref{eq7}) is more than 14 times larger than the
 corresponding standard deviation.  
By using this relation  we obtained $127\pm26$ for 
the amplitude of Solar Cycle~25. That is, we obtained  a  larger value than 
that $86\pm18$ was predicted earlier \citep{jj21}
 by using REL-I (non-revised)  and matches
 within the limits of the uncertainty  the  prediction made above 
from the revised REL-I.  It  also 
matches well  with the value $125 \pm 7$ obtained by using the amount of 
polar fields around the 
 minimum epoch of Solar Cycle~24 \citep{kumar22,jj23}. It is 
slightly larger than the amplitude of Solar Cycle~24 and is consistent with 
 the current trend of $SN_{\rm T}$.

\begin{figure}
\centering
\includegraphics[width=8.5cm]{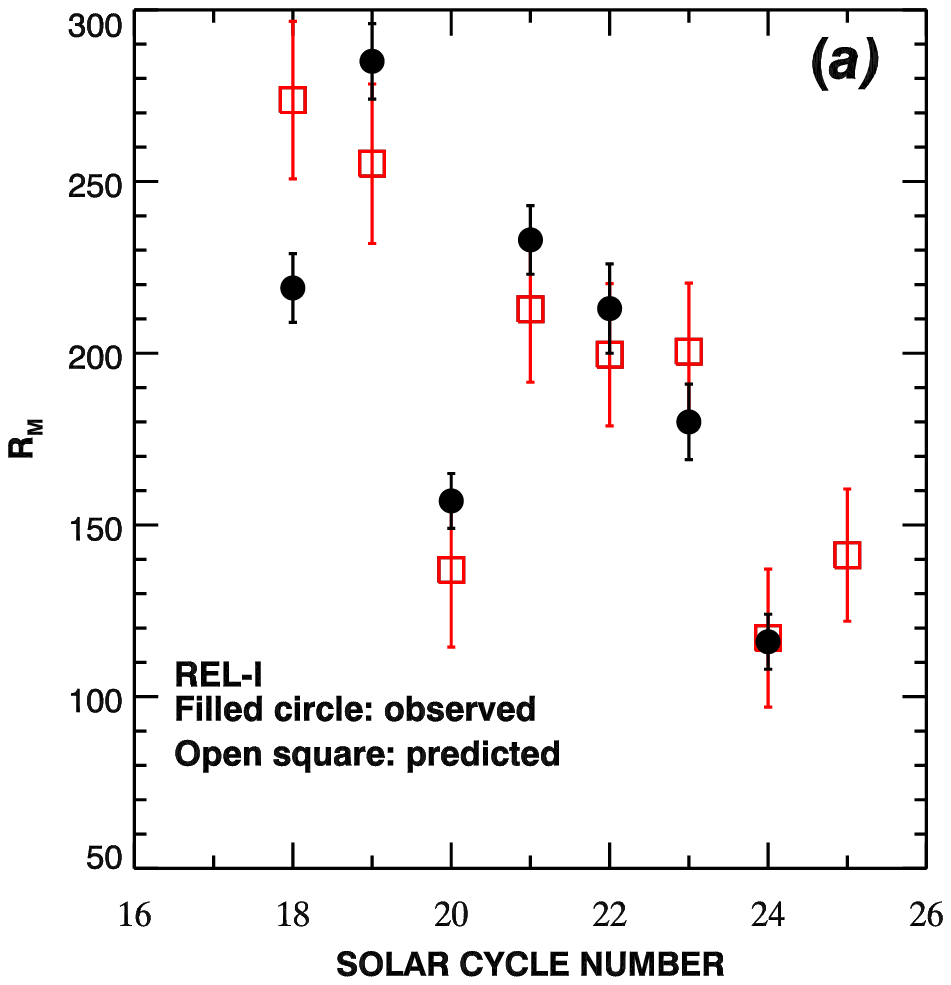}
\includegraphics[width=8.5cm]{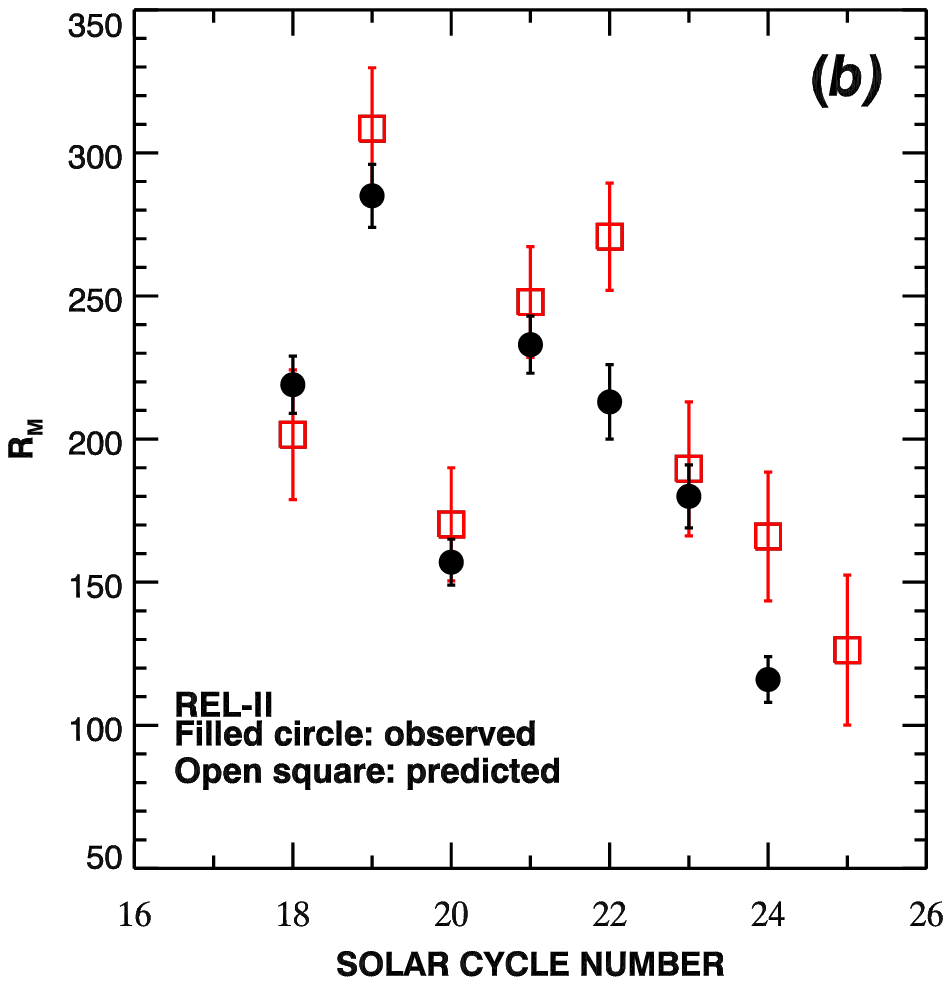}
\caption{Hindsight: Comparison of the observed  and the
predicted values of $R_{\rm M}$  of Solar Cycles~18\,--\,24.  
 The  predictions are made by using  
({\bf a}) REL-I and ({\bf b}) REL-II.
  The predicted values of $R_{\rm M}$  of Solar Cycle~25
are also shown.}
\label{f8}
\end{figure}

We checked the reliability  of the revised REL-I and REL-II.
Figs.~\ref{f8}a and \ref{f8}b show the observed values of $R_{\rm M}$ 
of Solar Cycles 18\,--\,24 and the predictions were made by using  
   REL-I and REL-II, respectively. In these figures 
we have also shown the predicted values of $R_{\rm M}$ of
 Solar Cycle 25.  
As can be seen in Fig.~\ref{f8}a except for Solar Cycle~18, 
remaining all solar cycles the predicted values closely agree with 
the corresponding observed values. We don't know why we obtained 
 the incorrect value for Solar Cycle~18 (any way poor statistics, 
 i.e. only 5 data points).
 There is also some concern about reliability of $T^*_{\rm M}$ because it 
is considerably small (about 5 months only). However, 
it contains a reasonably large amount of data (see Table~\ref{table3}).
 Therefore,  the consistency  of REL-I seems to be not bad.
   Hence, the prediction for Solar Cycle~25 that made by using REL-I seems
 to be reasonably reliable and we believe that 
by using this relation  a reasonable good prediction
 could also be made in future for an upcoming solar cycle.
As can be seen in Fig.~\ref{f8}b, the predicted values of Solar Cycles 22 and 24 
do not agree  with the corresponding observed values. Remaining all other 
cycles the agreement is good. Hence,  
the prediction for Solar Cycle 25 seems to be also reasonably reliable,  but  
 by using REL-II  a reasonable good prediction may be possible mostly only 
for an odd-numbered solar cycle.

\subsection{130-year periodicity in solar activity and similar solar cycles} 
The extrapolation of a cosine fit to the amplitudes of Solar Cycles 12\,--\,24 
  indicated that Solar Cycle~25 will be slightly larger than 
the weak Solar Cycle~24 \citep{jj22,jj23}. A  number of
 authors predicted  the Dalton minimum like low level of  
activity around  Solar Cycle~25 \citep[e.g.,][]{kom19,cob21}. 
\cite{du20a} found that Solar Cycles 24, 15, 12, 14, 17,  
and 10 (in that order) are most similar cycles to  Solar Cycle 25  
(note that this list is not containing Solar Cycle 13) and 
 predicted $137.8\pm 31.3$ for the amplitude of Solar Cycle 25,  which is 
 larger than the amplitude of Solar Cycle~24. 
The cosine fit to the amplitudes of Solar Cycles 12\,--\,24 suggested 
 that there exists a $\approx$130-year long-term cycle  in the amplitude
 modulation \citep{jj22,jj23}. In view of this and   the
 relative strengths of the amplitudes of solar cycles inferred from the
 patterns of the ratio $R_{\rm M}/R_A$ above, we can expect that the
 pattern of 
Solar Cycle~25 may be similar to that of  Solar Cycle 13. 

\begin{figure}
\centering
\includegraphics[width=8.5cm]{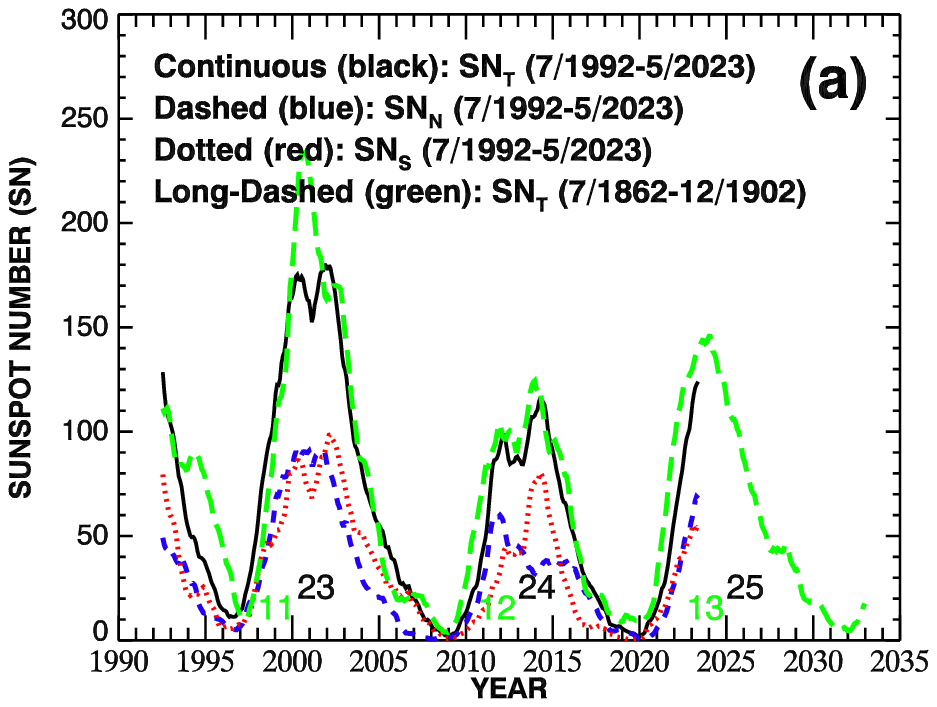}
\includegraphics[width=8.5cm]{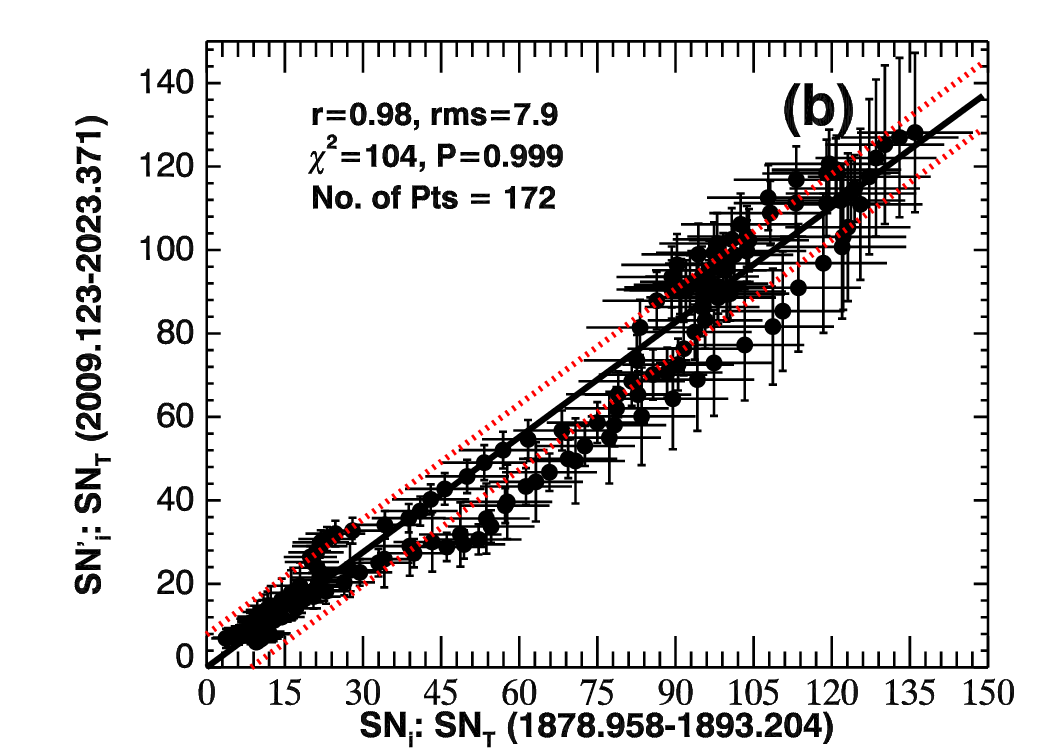}
\caption{({\bf a}) Variations in the 13-month smoothed monthly
mean   $SN_{\rm T}$, $SN_{\rm N}$, $SN_{\rm S}$,
during the period July/1992\,--\,May/2023. The corresponding 
Waldmeier solar cycles numbers are shown. Variations in the 13-month
 smoothed monthly mean $SN_{\rm T}$ during the period
  July/1862\,--\,December/1902 is also shown by forward  shifting  
 the epochs by 130-years.
 ({\bf b}) Correlation between the $SN_{\rm T}$ during the period 
1878.958\,--\,1893.204 and the $SN_{\rm T}$ during the period
 2009.123\,--\,2023.371. The horizontal and vertical error bars 
represent the  errors in $SN_i$ and $SN^\prime_i$, where $i =1$,\dots,172,
 i.e. the errors in the values  of   $SN_{\rm T}$ of the former and latter 
 periods, respectively. The {\it continuous line} represents the best-fit 
linear relation. The dotted curves are drawn at 1$\sigma$ level.
 The values of the correlation coefficient ($r$), $\chi^2$ 
and the corresponding probability ($P$), and the number of data points are
also shown.}
\label{f9}
\end{figure}

Fig.~\ref{f9}a shows the 
variations in the 13-month smoothed monthly
mean values of  $SN_{\rm T}$ and the corresponding values of 
northern hemisphere ($SN_{\rm N}$) and southern hemisphere ($SN_{\rm S}$),
during the period July/1992\,--\,May/2023 
(taken from  {\sf www.sidc.be/silso/datafiles}). 
 Variation in the 13-month
 smoothed monthly mean $SN_{\rm T}$ during the period
  July/1862\,--\,December/1902, i.e. this period 
included Solar Cycles 11, 12, and 13, 
is also shown by shifting the corresponding  epochs by 130-years (130-year
is added to the epochs). 
 As can be seen in this figure the shape of the
 whole Solar Cycle~24 and the rising phase of Solar Cycle 25 (for which the
data are available)  very closely resemble with the shape of Solar Cycle 12 
and the rising phase of Solar Cycle~13, respectively. However, 
 the  shapes of Solar Cycles 11 and 23 do not match each other well,
 particularly there exits no match between the maxima of these cycles.
 The variations in
 $SN_{\rm N}$ and $SN_{\rm S}$ during Solar Cycles 24  are  
 similar to the variations in the northern and southern hemispheres'  
13-month smoothed monthly mean areas of sunspot groups during Solar Cycle 12 
shown in Fig.~1 of  \cite{jj19,jj20}  and  also see Fig.~7 in \cite{veronig21}.
 Fig.~\ref{f9}b shows the  correlation between the $SN_{\rm T}$ 
during the period 1878.958\,--\,1893.204  (included full Solar Cycle~12
 and the rising phase of Solar Cycle 13)  and
 the $SN_{\rm T}$ during the period 2009.123\,--\,2023.371
 (included full Solar Cycle~24 and the rise phase 
of Solar Cycle 25).  We determined linear least-square-fit to 
these data.   To obtain insignificant $\chi^2$
the latter time series is shifted backward with respect to the 
former by two-months  (approximate).

 Let $SN_i$ is the value of  $SN_{\rm T}$  at an epoch $t_i$ 
in the interval 1878.958\,--\,1893.204 and 
  $SN^\prime_i$ is the value of  $SN_{\rm T}$  at an epoch $t^\prime_i$ 
in the interval 2009.123\,--\,2023.371,
 where $i = 1$,\dots,172,  then obviously 

\begin{equation}
 t^\prime_i = t_i + 130 +\frac{2}{12}, 
\label{eq8}
\end{equation}
and  we obtained the  linear relation:

\begin{equation}
SN^\prime_i  = SN_i (0.919 \pm 0.017).
\label{eq9}
\end{equation}

This linear equation is forced to pass through the origin because
 $SN_{\rm T}$ is never less than zero (the intercept was found to be
 $-4.34\pm 1.31$).
The   values of the correlation coefficient ($r$), $\chi^2$
and the corresponding probability ($P$), and the number of data points are
also shown in Fig.~\ref{f9}b. Uncertainties of both the abscissa and
 ordinate are 
taken care in the calculation of the least-square-fit.   
This linear relation (Eq.~(\ref{eq9})) is reasonably accurate, i.e.
the value of $r$ is high and the value of $\chi^2$ is reasonably small 
(the value of $P$ is high) and moreover the ratio  (equal to 54)
 of the slope to its standard division is very high.

\begin{figure}
\centering
\includegraphics[width=\textwidth]{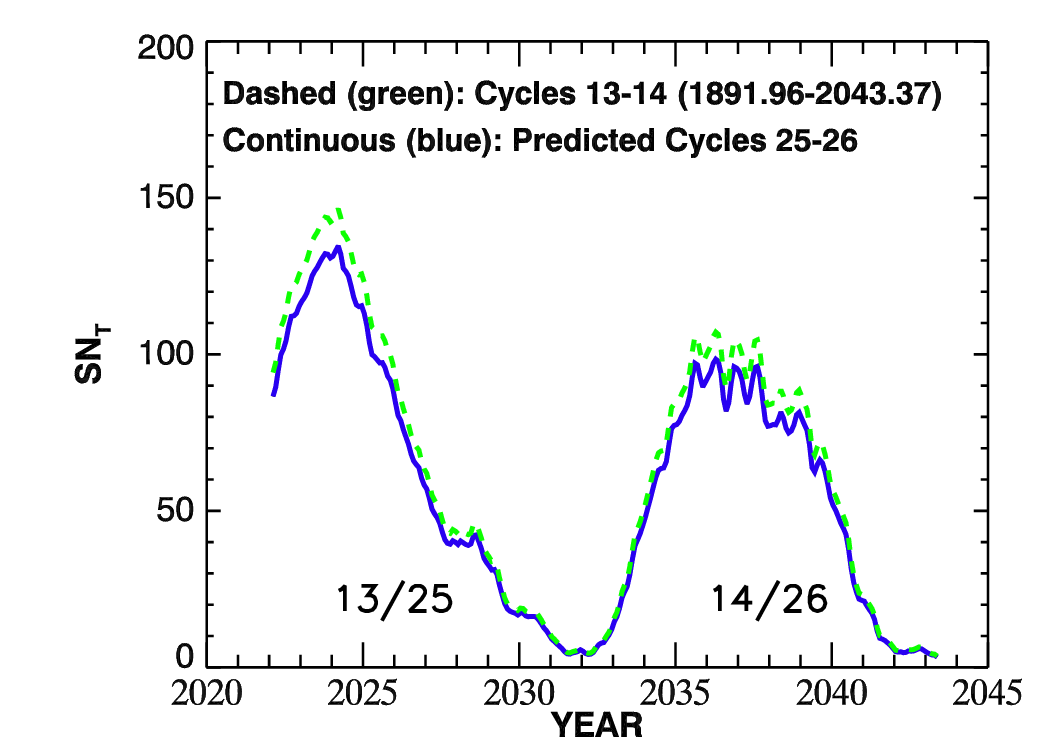}
\caption{Variations in the predicted 13-month smoothed monthly 
mean  $SN_{\rm T}$ during  Solar Cycles 25\,--\,26, obtained   by using 
 Eqs.~(\ref{eq8}) and (\ref{eq9}) and  the time series of 
13-month smoothed monthly mean  
$SN_{\rm T}$ during Solar Cycles 13\,--\,14. 
The variation in $SN_{\rm T}$ 
during Solar Cycles  13\,--\,14 is also shown.}   
\label{f10}
\end{figure}

Fig.~\ref{f10} shows  variations in the predicted 13-month smoothed monthly
mean  $SN^\prime_{\rm T}$ during Solar Cycles 25\,--\,26,  determined
  by using 
 Eqs.~(\ref{eq8}) and (\ref{eq9}) and the time series of 13-month smoothed 
monthly mean $SN_{\rm T}$ of  Solar Cycles 13\,--\,14. 
 In this figure the variation in 
the $SN_{\rm T}$ during Solar Cycles 13\,--\,14 is also shown. 
 As we can see in this figure  there exists a remarkable matching   
 between the  curves of the predicted  Solar Cycles 25\,--\,26
and Solar Cycles 13\,--\,14 in most of the times except during the maxima of 
the solar cycles. Obviously, the maximum and minimum epochs of 
the predicted Solar Cycles 25\,--\,26 are simply 
the values of the maximum and minimum epochs of Solar Cycles 13\,--\,14
 added   by 130-year and 2-months (Eq.~(\ref{eq8})). The amplitudes of the 
simulated Solar Cycles 25\,--\,26 are  to some extent lower (about 8\,\%) 
than the amplitudes of Solar Cycles 13\,--\,14. By using 
 the Eq.~(\ref{eq9})
it seems we are getting a  slight (about 8\,\%) underestimate for
 $SN^\prime_i$, i.e. for the predicted $SN_{\rm T}$. 
  Note that  
these simulations for Solar Cycles 25\,--\,26 are based on 
  about one-and-half solar cycles data (172 months) are used to 
derive Eq.~(\ref{eq9}). In principle by using Eqs.~(\ref{eq8}) and
 (\ref{eq9}) one can simulate the shapes of several upcoming solar cycles.   
 However, some more cycles 
data are required to make  better simulations/predictions for more  
solar cycles.   On the other hand, the predictions 
based on the extrapolated methods seem to be less impressive \citep{petro20}. 
 The shape of Solar Cycle 23 does
not match well with that of Solar Cycle 11. The 130-year period  of  sunspot 
activity was found from the  limited  (13 solar cycles) data. Although
it seems to be reasonably reliable because  the existence of a $130\pm 0.9$-year
 periodicity is found from a very large data (9400-year records) of $^{14}C$
 and $^{10}Be$ \citep{mcc13}, here we can make prediction 
for one solar cycle with a reasonable accuracy or tentatively at most two solar cycles. Therefore, in Fig.\ref{f10} 
 we have shown the predictions for only the two solar cycles, 25 and 26.
   We obtained $135 \pm 8$ for the 
amplitude of Solar Cycle~25  and  2024.21 (March 2024) for the 
corresponding maximum epoch.
 The end of this cycle may occur in 2032.21 (March 2032)
with the value $\approx$4 of  $SN_{\rm T}$ (minimum value of
 Solar Cycle~26). These epochs have  uncertainties by about  6-months 
(note that we have used 13-month smoothed monthly mean 
 $SN_{\rm T}$ time series).
 Solar Cycle~26 would be considerably smaller than Solar Cycle~25.  
Overall we can conclude that the shape of Solar Cycle 25 will be
 mostly similar to 
that of Solar Cycle 13.  The amplitude of Solar Cycle 25 will be considerably 
smaller than the average value 178.7 of the amplitudes of Solar 
Cycles~1\,--\,24 \citep{pesnell18} and the corresponding maximum 
 would occur in March/2024 ($\pm$ 6-month) and the Solar Cycle 25 will end
 in  March/2032 ($\pm$6-month). The length of Solar Cycle~25 will be about
 12-year, i.e. Solar Cycle~25 would be a long and below average size
 solar cycle and 
 will be followed by the weak Solar Cycle~26 (weakest in 12\,-\,13 decades). 

A number of authors predicted a weak-moderate Solar Cycle~25
but stronger than Solar Cycle~24 
\citep[e.g.,][]{cameron16,okoh18,ps18,petro18,du20a,du20b,du22,kakad20,
kumar21,kumar22,jj08,jj22,lu22,zhu22,braj22,nag23,luo24}. 
Our prediction here is consistent 
with these. Many authors predicted that maximum of Solar Cycle 25 will
 take place in 2025 \citep[e.g.,][]{ps18,okoh18,lab19,jj19}. However, it has
 been
 also  predicted that the maximum peak of this cycle will take place in 2024
 \citep{sb17,bn18,petro18,li18,du20a,ahl22,jas23,luo24}. 
 Recently, \cite{jha24}  in their advective flux transport model
assumed the activity during Solar Cycle 25 will be similar 
to  that of during Solar Cycle 13 and  predicted the peak of Solar Cycle~25
 will occur between April and August of 2024. Our prediction here is almost 
the same as this. 
 \cite{lu22} by using a  bi-modal forecasting model 
 predicted that Solar Cycle~25 will be single-peak structure and the peak
 will occur in October/2024. From a method of similar cycles \cite{du20a} has
 predicted that a secondary peak in Solar Cycle~25 will occur eight-months
 earlier than the major peak.   
   As in the case of Solar Cycle~13 \cite[e.g.,][]{jj23},
 in Solar Cycle 25 the second peak may be the main peak and would occur
about one-year after the first peak (secondary peak), but the ratio of the
 value of the first peak  to that of the
 second peak   would be large ($\approx$0.95).     
 The first peak of Solar Cycle 25  might 
have been already occurred and could not be clearly seen in the time series
 of 13-month smoothed monthly mean $SN_{\rm T}$ (on the other hand
the current level of the rising phase may represent the first peak).

\section{Discussion and conclusions} 
Here we analysed the combined data of sunspot groups from GPR
during the period 1874\,--\,1976 and DPD during the period 1977\,--\,2017 and
determined the  monthly mean, annual mean, and 13-month smoothed monthly
 mean whole sphere  sunspot-group area
 (WSGA).  We have also analysed the monthly mean, annual mean, and       
 13-month smoothed   
 monthly mean version 2 of international sunspot number ($SN_{\rm T}$)
 during the period 1874\,--\,2017.
 We determined correlation between  WSGA and $SN_{\rm T}$ and 
found that the statistical significance 
of the correlation is  good in each solar cycle.
  We fitted the
  annual mean  WSGA and $SN_{\rm T}$ data during each of
 Solar Cycles~12\,--\,24 separately to the linear and nonlinear
 (parabola)
 forms. In the cases of Solar Cycles 14, 17, and 24 the nonlinear fits
are found better than the corresponding linear fits.
  We find that there exists a secular
 decreasing trend in the slope   of the
 WSGA--$SN_{\rm T}$ linear relation during Solar Cycles~12\,--\,24,
superimposed on it  a weak  $\approx77$-year variation. 
The slope of the average Solar Cycle~15 is found to be largest 
and is at about 95\,\% confidence level. 
  A secular decreasing trend is also seen in the coefficient of the
 first order term of the nonlinear relation.
There is no significant correlation  between the slope/coefficient and the  
amplitude of a solar cycle. The  cycle-to-cycle variation 
in the ratio of the amplitude ($R_{\rm M}$) of a solar cycle to the value
 ($R_A$) of WSGA at the maximum epoch of the solar cycle  is found to be very 
 during Solar Cycles~12\,--\,24 and closely 
similar to that of the slope. The existence of $\approx77$-year
long-term variation in this ratio $R_{\rm M}/R_A$ is seen  clearly 
(amplitude is equal to  $\approx2\sigma$).
  Based on this long-term variation of the  ratio we
 predicted that: Solar Cycle 25 is
 larger than the small Solar Cycle 24 but its amplitude would be 
smaller than the average value of the amplitudes of Solar Cycles~1\,--\,24, 
 Solar Cycle 26 would be smaller than Solar Cycle 25, and   
the solar cycle pair (26, 27) will satisfy the Gnevyshev-Ohl rule or G-O rule
 of solar cycles, and Solar Cycle 28 will be smaller than Solar Cycle 27. 
 Using an  our earlier method (now slightly revised), i.e. using
 high correlations of 
the amplitude of a solar cycle with the sums of the areas of sunspot groups
in $0$\,--\,$10^\circ$ latitude intervals of the northern hemisphere
 during  3.75-year interval around the minimum--and 
the southern hemisphere during 0.4-year   
near the maximum--of the corresponding preceding solar cycle, we predicted the 
 values  $127 \pm 26$ and $141 \pm 19$ for the amplitude of
 Solar Cycle~25, respectively.
Based on the existence of $\approx$130-year periodicity  in the cycle-to-cycle
variation of the amplitudes of Solar Cycles 12\,--\,24
\citep{jj22,jj23} we find the shape of Solar Cycle~25 would be similar 
to that of Solar Cycle~13 and predicted $135 \pm 8$ for the
amplitude of Solar Cycle~25,  2024.21 (March 2024)$\pm$6-month  for the
corresponding maximum epoch, and ending in 2032.21
 (March 2032)$\pm$ 6-month with the value $\approx 4$ of  $SN_{\rm T}$ 
(minimum value of Solar Cycle~26). 
 The length of Solar Cycle 25 is expected to $\approx$12-year.
 The Solar Cycle~26 would be a small 
cycle and  is expected to be similar as that of Solar Cycle~14.  

The  ratio $R_{\rm M}/R_A$ is large  in some cycles (15, 21) and small 
in some other cycles.  It indicates that in the former case the ratio of
 the  number of large size sunspot groups to that of the small size 
sunspot groups may be small, whereas in the latter case this ratio  may be
 large. In the former case while the magnetic structures of
 sunspot groups rising through the convection zone disintegrate or fragment into more small structures \citep[e.g.,][]{gj02} than in the latter case.

Because of variation in the slopes of the 
linear relationships of WSGA and $SN_{\rm T}$ of different solar cycles,  
this analysis also suggest that the prediction made by using the slope of the 
relationship between WSGA and $SN_{\rm T}$  determined from the data of whole
 period 1874\,--\,2017 might have  a large uncertainty \citep{jj21}. 
 The corresponding prediction made by using the linear
relationship between the logarithm values of WSGA and $SN_{\rm T}$ 
 \citep{jj23} may be also to some extent erroneous
because in the present analysis  we also find that  
the linear-least-square best fits of the  logarithm values 
of WSGA and $SN_{\rm T}$ of individual solar cycles are   not well defined.
The fit of the logarithm values of WSGA and $SN_{\rm T}$ of  the
 entire period of 1874\,--\,2017 may be better than the corresponding 
linear fit of the original values due to the ratio of the number of small size
 sunspot groups to the number of large size sunspot groups is not the same 
in all solar cycles \citep{jj16}, may be considerably large in some solar
 cycles.

A reasonable correlation exists  between solar cycle maximum ($R_{\rm M}$)
and the preceding minimum \citep{tlatov09,hath15}. But it is not possible
 to make 
a precise prediction for the former from the latter~\citep{du10}. However,
 it seems possible from the 13-month smoothed minimum just after
onset of the cycle \citep{ramesh00,ramesh12}. From this method
 it seems possible 
to predict $R_{\rm M}$ of a solar cycle by about three years in advance.
 From our  method, i.e. by using the relation REL-I,  the $R_{\rm M}$ 
of  a solar cycle may be possible to  
 predict by nine years in advance and by using REL-II the $R_{\rm M}$ of
mostly only an odd-numbered solar cycle  may be possible to predict by 
 about thirteen years in advance.
It may be worth to  note here that the two main ingredients of solar 
dynamo  mechanism,  deferential rotation and meridional flow,  differ  in
 odd- and even-numbered solar cycles \citep{jj03,jj08,jbu05}.
 As we have already
 discussed in \cite{jj08}, equatorial crossing of magnetic flux due to  
rotation of the Sun on inclined axis could be the
 reason behind the relations REL-I and REL-II. 

Regarding  the existence of  $\approx$130-year periodicity in the amplitude 
modulation of solar cycles \citep{jj22,jj23} that we have used here 
for finding  the similar solar cycles and predicting the repetitions of the 
patterns of Solar Cycles  12\,--\,14 in Solar Cycles 24\,--\,26,  
the existence of $\approx$130-year periodicity in some solar activity 
related phenomena are known. \cite{att90a,att90b} found the existence of 
 $\approx$22-year, $\approx$88-year (Gleissberg cycle), and
 $\approx$132-year periodicities  in aurorae, $^{14}C$ from 
tree rings, and $^{10}Be$  from polar ice and suggested that 
1/88-year and 1/132-year frequencies might be two sub-harmonics of the 
 22-year Hale cycle. According to them these results support the idea that 
the Sun behaves as nonlinear system forced internally by
 22-year torsional 
magnetohydrodynamic oscillation \citep{brace86,brace88,gjet92,gj92,gj95}. 
The required forcing  may be also coming
 from the alignments of Venus, Earth, and Jupiter.
 These  are the tidally dominant
planets and whose alignments might have a significant role in the solar cycle  
 mechanism \citep[][and references therein]{wood72,irgw13,stef21}.
 \cite{mcc13} found the existence of a $130\pm 0.9$-year periodicity
 in 9400-year records of $^{14}C$ and $^{10}Be$. It may be worth to note 
that $213V=131E=11J=131$-year,
 where $V = 0.615$-year, $E = 1 $-year, and $J = 11.9$-year, are the orbital
 periods of Venus, Earth, and Jupiter, respectively. That is,  
the relative orbital positions (configurations) of these planets vary 
differently during each of the 12 solar cycles that occur within
 the 131-year period  and the configurations would be similar in every
 13th solar cycle. Hence, the  influences of these planets on solar activity  
 could be similar during every 13th solar cycle. This may be a 
reason for the shape of Solar Cycle 24 is similar 
to that of Solar Cycle 12 and the shape of Solar Cycle 25 is expected to 
 be similar to that of Solar Cycle~13.  
The basic process involved  may be  the Sun's spin-orbit coupling
 \citep{gj95,juckett00,juckett03,jj03,jj05}.  
Beside the configurations of the giant planets, some specific 
alignments of other planets with 
the giant planets may also 
  have a role in the Sun's spin-orbit coupling \citep{ww65,irgw13,stef21}.
However, the role of the orbital motions of the planets in solar dynamo
 is not yet clear \citep{charb22}.

\vspace{0.5cm}
 \noindent{\large \bf Acknowledgments}

\vspace{0.3cm}
 The author thanks the anonymous reviewers for very useful comments 
and suggestions. The author acknowledges the work of all the
 people who contribute  and maintain the GPR and DPD  sunspot databases.
The sunspot-number data are provided by WDC-SILSO, Royal Observatory of
Belgium, Brussels.


{}
\clearpage

\clearpage

\begin{table}
{\scriptsize
\caption[]{Values of the slope ($m$) and its uncertainty ($\sigma_m$)
 of the best fit linear relation (Eq.~(\ref{eq1})) of the annual mean 
values of WSGA and $SN_{\rm T}$ 
of a solar cycle (SC).  The values of $\chi^2$ and the corresponding 
probability ($P$), the ratio $m/\sigma_m$, correlation coefficient 
($\rm r$), and the number of data points (N) are  given. The 
symbol $^{\mathrm a}$ indicates the incomplete data of Solar Cycle~24.
The values of the amplitude ($R_{\rm M}$) of a solar cycle and the 
 value of ($R_A$) of WSGA at the epoch of $R_{\rm M}$ and the corresponding
uncertainties 
$\sigma_R$ and  $\sigma_A$,   respectively, are also given.}
\smallskip
\begin{tabular}{lccccccccccccccccccccccccccccccccc}
\hline
SC &$R_{\rm M}$&$\sigma_R$&$R_A$&$\sigma_A$&$m$ &$\sigma_m$ & $m/\sigma_m$& $\chi^2$ & $P$& $\rm r$&  N\\
\hline
 12& 124.4& 12.5&1370.7&121.7& 0.101&  0.006&  16.83&   5.27&   0.87&   0.99&  12\\
 13& 146.5&10.8 &1616.0&109.9&  0.099&  0.005&  19.80&   9.73&   0.46&   0.99&  12\\
 14&107.1&9.2&1043.9& 139.5& 0.101&  0.005&  20.20&  11.77&   0.23&   0.97&  11\\
 15&175.7&11.8&1535.4&170.5&  0.111&  0.005&  22.20&   7.06&   0.53&   0.99&  10\\
 16&130.2&10.2&1324.0&122.9&  0.089&  0.005&  17.80&   6.57&   0.58&   0.99&  10\\
 17&198.6&12.6&2119.5&175.7&  0.097&  0.004&  24.25&  11.51&   0.24&   0.99&  11\\
 18&218.7&10.3&2641.4&209.7&  0.084&  0.003&  28.00&   9.62&   0.29&   0.99&  10\\
 19&285.0&11.3&3441.4&208.2&  0.085&  0.002&  42.50&  24.88&   0.00&   0.99&  10\\
 20&156.6& 8.4&1556.3& 81.8&  0.088&  0.003&  29.33&   6.84&   0.74&   0.99&  12\\
 21&232.9&10.2&2121.2&161.9&  0.098&  0.003&  32.67&  16.94&   0.03&   0.98&  10\\
 22&212.5&12.7&2268.7&193.4&  0.084&  0.003&  28.00&   8.51&   0.39&   0.99&  10\\
 23&180.3&10.8&2157.3&205.6&  0.083&  0.003&  27.67&   6.92&   0.73& $\approx1$&  12\\
 24$^{\mathrm a}$&116.4&8.2&1599.8&115.9& 0.081&0.004&20.25&14.83&0.06&0.97&  10\\
 Whole&&&&&  0.090&  0.001&  90.00& 249.80&   0.00&   0.98& 144\\
\hline
\end{tabular}
\label{table1}
}
\end{table}

\begin{table}
{\scriptsize
\caption[]{Values of the coefficients ($m_1$ and $m_2$) and their
 uncertainties ($\sigma_{m_1}$ and $\sigma_{m_2}$)
 of the best fit nonlinear relation  (Eq.~(\ref{eq2})) of the annual mean 
values of  WSGA and $SN_{\rm T}$ of a solar cycle (SC).
 The values of  ratios $m_1/\sigma_{m_1}$ and $m_2/\sigma_{m_2}$ and 
 the values of $\chi^2$ and the corresponding probability ($P$) are also given.
 The symbol $^{\mathrm a}$ indicates the incomplete data of Solar Cycle~24.}
\smallskip
\begin{tabular}{lccccccccccccccccccccccccccccccccc}
\hline
SC &$m_1$&$\sigma_{m_1}$ &$m_2$&$\sigma_{m_2}$ & $m_1/\sigma_{m_1}$& $m_2/\sigma_{m_2}$& $\chi^2$ & $P$\\
\hline
 12&  0.129&  0.021& $-2.90\times10^{-5}$&$ 2.12\times10^{-5}$&   6.14&$  -1.37$&  13.40&   0.20\\
 13&  0.105&  0.014& $-5.53\times10^{-6}$&$ 1.10\times10^{-5}$&   7.50&$  -0.50$&  23.23&   0.01\\
 14&  0.146&  0.016& $-4.71\times10^{-5}$&$ 1.53\times10^{-5}$&   9.12&$  -3.08$&   3.42&   0.97\\
 15&  0.102&  0.012& $ 8.25\times10^{-6}$&$ 1.03\times10^{-5}$&   8.50&$   0.80$&   6.74&   0.57\\
 16&  0.106&  0.018& $-1.39\times10^{-5}$&$ 1.38\times10^{-5}$&   5.89&$  -1.01$&  10.42&   0.32\\
 17&  0.124&  0.012& $-1.53\times10^{-5}$&$ 6.19\times10^{-6}$&  10.33&$  -2.47$&   5.99&   0.82\\
 18&  0.098&  0.009& $-6.31\times10^{-6}$&$ 4.01\times10^{-6}$&  10.89&$  -1.57$&  14.50&   0.11\\
 19&  0.109&  0.007& $-9.09\times10^{-6}$&$ 2.42\times10^{-6}$&  15.57&$  -3.76$&   9.66&   0.29\\
 20&  0.100&  0.011& $-8.12\times10^{-6}$&$ 6.87\times10^{-6}$&   9.09&$  -1.18$&  43.41&   0.00\\
 21&  0.093&  0.012& $ 2.13\times10^{-6}$&$ 5.81\times10^{-6}$&   7.75&$   0.37$&  19.85&   0.02\\
 22&  0.104&  0.010& $-9.00\times10^{-6}$&$ 4.48\times10^{-6}$&  10.40&$  -2.01$&  11.71&   0.23\\
 23&  0.087&  0.009& $-1.96\times10^{-6}$&$ 4.70\times10^{-6}$&   9.66&$  -0.42$&   8.10&   0.70\\
 24$^{\mathrm a}$&  0.111&  0.009& $-2.51\times10^{-5}$&$6.97\times10^{-6}$&12.33&$-3.60$&4.16&0.84\\
 Whole&  0.103&  0.002&$-7.14\times10^{-6}$&$1.04\times10^{-6}$&51.50&$-6.86$ & 296.88&   0.00\\
\hline
\end{tabular}
\label{table2}
}
\end{table}

\begin{table}
{\scriptsize
\centering
\caption{$T_{\rm M}$ and $T_{\rm m}$ are  the maximum and minimum epochs 
of a solar cycle ($n$). 
 $A^*_{\rm M}$ and $A^*_{\rm m}$ represent  the sums of the areas (msh) of the 
sunspot groups (normalized by 1000) in  $0^\circ-10^\circ$ latitude 
intervals of the southern  and northern hemispheres
  during  the time intervals $T^*_{\rm M} = 0.4$-year near 
 $T_{\rm M}$
 (i.e.,  $T^*_{\rm M} +1.2$-year\,--\,$T^*_{\rm M} + 1.6$-year) and 
 $T^*_{\rm m} = 3.75$-year around 
 $T_{\rm m}$ (i.e.,  $T_{\rm m} - 1.4$-year\,--\,$T_{\rm m} + 2.35$-year) 
  of a solar cycle, respectively. The symbol $^{\mathrm b}$ indicates 
 that the epochs of the first peaks are used (see the text).}

\smallskip
\begin{tabular}{lcccccccccc}
\hline
  \noalign{\smallskip}
$n$&$T_{\rm M}$&$T^*_{\rm M}$&$A^*_{\rm M}$ & $T_{\rm m}$&$T^*_{\rm m}$&$A^*_{\rm m}$\\
  \noalign{\smallskip}
\hline
  \noalign{\smallskip}
12&1883.96&1885.16-1885.56&  19.75&1878.96&1877.56-1881.31&  10.86\\
13&1894.04&1895.24-1895.64&  19.00&1890.20&1888.80-1892.55&  11.62\\
14&1906.12&1907.32-1907.72&  25.76&1902.04&1900.64-1904.39&  11.77\\
15&1917.62&1918.82-1919.22&  28.19&1913.62&1912.22-1915.97&   9.79\\
16&1928.29&1929.49-1929.89&  33.07&1923.62&1922.22-1925.97&  36.47\\
17&1937.29&1938.49-1938.89&  54.28&1933.71&1932.31-1936.06&  37.48\\
18&1947.37&1948.57-1948.97&  65.21&1944.12&1942.72-1946.47&  69.91\\
19&1958.20&1959.40-1959.80&  22.78&1954.29&1952.89-1956.64&  23.93\\
20&1968.87&1970.07-1970.47&  45.93&1964.79&1963.39-1967.14&  54.42\\
21&1979.96&1981.16-1981.56&  40.56&1976.21&1974.81-1978.56&  65.30\\
22&1989.87&1991.07-1991.47&  40.52&1986.71&1985.31-1989.06&  35.06\\
23&2000.29$^{\mathrm b}$&2001.49-2001.89&14.55&1996.62&1995.22-1998.97&  24.84\\
24&2012.21$^{\mathrm b}$&2013.41-2013.81&22.33&2008.96&2007.56-2011.31&  10.93\\
\hline
  \noalign{\smallskip}
\end{tabular}
\label{table3}
}
\end{table}

\begin{thebibliography}{}
\bibitem[\protect\citeauthoryear{Ahluwalia}{2022}]{ahl22}
Ahluwalia, H.S., 2022. Forecast for sunspot cycle 25 activity. \adv\ 69, 794.
\doiurl {10.1016/j.asr.2021.09.035}.
\bibitem[\protect\citeauthoryear{Attolini \etal}{1990a}]{att90a}
Attolini, M.R., Cecchini, S., Galli, M., Nanni, T., 1990a.
 The subharmonics of the  22-year solar cycle. {\it NCimC.} 13, 131.
\doiurl{10.1007/BF02515782}.
\bibitem[\protect\citeauthoryear{Attolini \etal}{1990b}]{att90b}
Attolini, M.R., Cecchini, S., Galli, M., Nanni, T., 1990b. 
 On the persistence of the 22-YEAR  solar cycle. \solphys\ 125, 389.
\doiurl{10.1007/BF00158414}.
\bibitem[\protect\citeauthoryear{Bhowmik}{2018}]{bn18}
Bhowmik, P., Nandy, D., 2018.  Prediction of the strength and timing of 
sunspot cycle 25 reveal decadal-scale space environmental conditions.
\Natco\ 9, 5209.
 \doiurl{10.1038/s41467-018-07690-0}.
\bibitem[\protect\citeauthoryear{Bracewell}{1986}]{brace86}
Bracewell, R.N., 1986. Simulating the sunspot cycle. \nat\ 323, 516.
  \doiurl{10.1038/323516a0}.
\bibitem[\protect\citeauthoryear{Bracewell}{1988}]{brace88}
Bracewell, R.N., 1988. Three-halves law in sunspot cycle shape.
 \mnras\ 230, 535.    \doiurl{10.1093/230.4.535}.
\bibitem[\protect\citeauthoryear{Braj\v{s}a \etal}{2022}]{braj22}
Braj\v{s}a. R., Verbanac, G., Bandi\'{c}, M., Hanslmeier, A., Skoki\'{c}, I., 
Sudar, D., 2022. A prediction for the 25th solar cycle maximum amplitude.
{\it AN}, 343, e13960.
  \doiurl{10.1002/asna.202113960}.
\bibitem[\protect\citeauthoryear{Cameron \etal}{2016}]{cameron16}
Cameron, R.H., Jiang, J., Sch{\"u}ssler, M., 2016. Solar cycle 25: Another moderate cycle? \apjl\ 823, l22.
  \doiurl{10.3847/2041-8205/823/2/L22}.
\bibitem[\protect\citeauthoryear{Charbonneau}{2022}]{charb22}
Charbonneau, P., 2022. External Forcing of the Solar Dynamo. 
{\it Front. Astron. Space Sci.} 9, 853676.
\doiurl{10.3389/fspas.2022.853676}. 
\bibitem[\protect\citeauthoryear{Clette and Lef\'evre}{2016}]{clette16}
Clette, F.,  Lef\'evre, L., 2016. The new sunspot number: Assembling all 
corrections.  \solphys\ 291, 2629.\doiurl{10.1007/s11207-016-1014-y}.
\bibitem[\protect\citeauthoryear{Coban \etal}{2021}]{cob21}
Coban, G.C., Raheem, A., Cavus, H., Asghan-Targhi, M., 2021.
Can solar cycle 25 be a new Dalton minimum? \solphys\ 296, 156.
 \doiurl{10.1007/s11207-021-01906-1}
\bibitem[\protect\citeauthoryear{Dikpati, \etal}{2008}]{dgt08}
Dikpati, M., Gilman, P. A.,  de Toma, G., 2008. The Waldmeier effect: An 
artifact of the definition of wolf sunspot number? \apjl\  673, L99.
  \doiurl{10.1086/527360}.
\bibitem[\protect\citeauthoryear{{Du}}{2020a}]{du20a}
Du, Z., 2020a. Predicting the shape of solar cycle 25 using a
similar-cycle method. \solphys\ 295, 134.
\doiurl{10.1007/s11207-020-01701-4}.
\bibitem[\protect\citeauthoryear{{Du}}{2020b}]{du20b}
Du, Z., 2020b. Predicting the amplitude of  solar cycles 25 using 
the value 39 months before the solar minimum. \solphys\ 295, 147.
\doiurl{10.1007/s11207-020-01720-1}.
\bibitem[\protect\citeauthoryear{{Du}}{2022}]{du22}
Du, Z., 2022. Predicting the amplitude of  solar cycle 25 using the early value 
of the rising phase. \solphys\ 297, 61.
 \doiurl{10.1007/s11207-022-01991-w}.
\bibitem[\protect\citeauthoryear{{Du and Wang}}{2010}]{du10}
Du, Z.L., Wang, H.N., 2010.  Does a low solar cycle minimum hint at
 a weak upcoming cycle? {\it Res. in Astron. and Astrophys.} 10, 950.
 \doiurl{10.1088/1674-4527/10/10/002}.
\bibitem[\protect\citeauthoryear{Gnevyshev and Ohl}{1948}]{go48}
Gnevyshev, M.N., Ohl, A.I., 1948. About 22-year cycle of solar activity. 
{\it AZh} 25, 18.
\bibitem[\protect\citeauthoryear{Gokhale and Javaraiah}{1992}]{gj92}
 Gokhale, M.H., Javaraiah, J., 1992. Global modes constituting the solar
 magnetic cycle - Part Two. \solphys\ 138, 399.
 \doiurl{10.1007/BF00151923}.
\bibitem[\protect\citeauthoryear{Gokhale and Javaraiah}{1995}]{gj95}
 Gokhale, M.H., Javaraiah, J., 1995. Global modes constituting the solar
 magnetic cycle - Part Three. \solphys\ 156, 157.
 \doiurl{10.1007/BF00669582}.
\bibitem[\protect\citeauthoryear{Gokhale and Javaraiah}{2002}]{gj02}
 Gokhale, M.H., Javaraiah, J.  in J. Javaraiah and M.H. Gokhale (eds),
 The Sun's rotation, Nova Science, New York, pp.109, 2002.
\bibitem[\protect\citeauthoryear{Gokhale \etal}{1992}]{gjet92}
 Gokhale, M.H., Javaraiah, J. Narayanakutty, K., Varghese, B.A., 1992.
 Global modes constituting the solar magnetic cycle - Part One.
 \solphys\ 138, 35. \doiurl{10.1007/BF00146195}.
\bibitem[\protect\citeauthoryear{Hathaway}{2015}]{hath15}
Hathaway, D.H., 2015. The Solar Cycle. {\it Liv. Rev. Solar. Phys.} 12, 4.
 \doiurl{10.1007/1rsp-2015-4}.
\bibitem[\protect\citeauthoryear{Hathaway \etal}{2002}]{hath02}
Hathaway, D.H., Wilson, R.M., Reichmann, E.J., 2002. Group sunspot numbers;
 sunspot cycle characteristics. \solphys\ 211, 357.
 \doiurl{10.1023/A:1022425402664}.
\bibitem[\protect\citeauthoryear{Jaswal, \etal}{2023}]{jas23}
Jaswal, P., Saha, C., Nandy, D., 2023. Discovery of a relation between
 the decay rate of the Sun's magnetic dipole and the growth rate of
 the following sunspot cycle: a new precursor
 for solar cycle prediction. \mnras\  528, L27.
  \doiurl{10.1093/mnrasl/slad122}.
\bibitem[\protect\citeauthoryear{Javaraiah}{2003}]{jj03}
Javaraiah, J., 2003. Long--term variations in the solar differential rotation.
 \solphys\  212, 23. DOI: \doiurl{10.1023/A:1022912430585}.
\bibitem[\protect\citeauthoryear{Javaraiah}{2005}]{jj05}
Javaraiah, J., 2005. Sun's retrograde motion and violation of even-odd
 cycle rule in sunspot activity. \mnras\ 362, 1311.
  \doiurl{10.1111/j.1365-2966.2005.09403.x}.
\bibitem[\protect\citeauthoryear{Javaraiah}{2007}]{jj07}
Javaraiah, J., 2007. North--south asymmetry in solar activity: predicting 
the amplitude of the next solar cycle. \mnras\ 377, L34.
  \doiurl{10.1111/j.1745-3933.2007.00298.x}.
\bibitem[\protect\citeauthoryear{Javaraiah}{2008}]{jj08}
Javaraiah, J., 2008. Predicting the amplitude of a solar cycle using
the north--south asymmetry in the previous cycle:
II. An improved prediction for solar cycle 24. \solphys\  252, 419.
  \doiurl{10.1007/s11207-008-9269-6}.
\bibitem[\protect\citeauthoryear{Javaraiah}{2012}]{jj12}
Javaraiah, J., 2012. The G--O rule and Waldmeier effect in the variations
of the numbers of large and small sunspot groups. \solphys\ 281, 827. 
  \doiurl{10.1007/s11207-012-0106-6}.
\bibitem[\protect\citeauthoryear{Javaraiah}{2015}]{jj15}
Javaraiah, J., 2015. Long--term variations in the north--south asymmetry
 of solar activity and solar cycle prediction, III: Prediction for the
 amplitude of solar cycle 25. \na\  34, 54. 
 \doiurl{10.1016/j.newast.2014.04.001}.
\bibitem[\protect\citeauthoryear{Javaraiah}{2016}]{jj16}
Javaraiah, J., 2016. North-south asymmetry in small and large sunspot group 
activity and violation of even-odd solar cycle rule. \apss\  361, 208.
 \doiurl{10.1007/s10509-016-2797-x}.
\bibitem[\protect\citeauthoryear{Javaraiah}{2017}]{jj17}
Javaraiah, J., 2017. Will solar cycles 25 and 26 be weaker than cycle 24?
\solphys\ 292, 172.
 \doiurl{10.1007/s11207-017-1197-x}.
\bibitem[\protect\citeauthoryear{Javaraiah}{2019}]{jj19}
Javaraiah, J., 2019. North--south asymmetry in solar activity and solar
cycle prediction, IV: Prediction for lengths of upcoming
solar cycles. \solphys\ 294, 64.  \doiurl{10.1007/s11207-019-1442-6}.
\bibitem[\protect\citeauthoryear{Javaraiah}{2020}]{jj20}
Javaraiah, J., 2020. Long--term periodicities in north--south asymmetry
of solar activity and alignments of the giant planets. \solphys\
  295, 8.  \doiurl{10.1007/s11207-019-1575-7}.
\bibitem[\protect\citeauthoryear{Javaraiah}{2021}]{jj21}
Javaraiah, J. 2021. North--south asymmetry in solar activity 
and solar cycle prediction, V: Prediction for the north--south asymmetry
 in the amplitude of Solar Cycle 25. \apss\  366, 16.
 \doiurl{10.1007/s10509-021-03922-w}.
\bibitem[\protect\citeauthoryear{Javaraiah}{2022}]{jj22}
Javaraiah, J., 2022. Long-term variations in solar activity: Predictions for
amplitude and north--south asymmetry of solar cycle 25.
 \solphys\ 297, 33.
 \doiurl{10.1007/s11207-022-01956-z}.
\bibitem[\protect\citeauthoryear{Javaraiah}{2023}]{jj23}
Javaraiah, J., 2023. Prediction for the amplitude and second maximum
 of Solar Cycle 25 and a comparison of the predictions based on
 strength of polar magnetic field and low-latitude sunspot area.
 \mnras\ 520, 5586.  \doiurl{10.1093/mnrasl/stad479}.
\bibitem[\protect\citeauthoryear{Javaraiah \etal}{2005}]{jbu05}
Javaraiah, J., Bertello, L., Ulrich, R.K., 2005. An interpretation of the 
differences in the solar differential rotation during even and odd sunspot 
cycles. \apj\ 626, 579. \doiurl{10.1086/429898}.
\bibitem[\protect\citeauthoryear{Jha and Upton}{2024}]{jha24}
Jha, B.K., Upton, L.A., 2024. Predicting the timing of the Solar Cycle 25 
polar field reversal. \apjl\ 962, L15. \doiurl{10.3847/2041-8213/ad20d2}.
\bibitem[\protect\citeauthoryear{Juckett}{2000}]{juckett00}
Juckett, D.A., 2000. Solar activity cycles, North/South asymmetries, and
 differential rotation associated with
solar spin-orbit variations. \solphys\ 191, 201.
 \doiurl{10.1023/A:1005226724316}.
\bibitem[\protect\citeauthoryear{Juckett}{2003}]{juckett03}
Juckett, D., 2003. Temporal variations of low-order spherical harmonic
representations of sunspot group patterns: Evidence for 
solar spin-orbit coupling. \aap\ 399, 731. 
 \doiurl{10.1051/0004-6361:20021923}.
\bibitem[\protect\citeauthoryear{Kakad \etal}{2020}]{kakad20}
Kakad, B., Kumar, R., Kakad, A., 2020. Randomness in sunspot number:
 A clue to predict solar cycle 25. \solphys\ 295, 88.
 \doiurl{10.1007/s11207-020-01655-7}.
\bibitem[\protect\citeauthoryear{Karak and Choudhuri}{2011}]{kc11}
Karak, B.B., Choudhuri, A.R., 2011. The Waldmeier effect and the flux
 transport solar dynamo. \mnras\ 410, 1503.
 \doiurl{10.1111/j.1365-2966.2010.17531.x}.
\bibitem[\protect\citeauthoryear{Kilcik \etal}{2011}]{kilc11}
Kilcik, A., Yurchyshyn, V.B., Abramenko, V., Goode, P.R.,
 Ozguc, A., Rozelot, J.P., Cao, W., 2011. Time distribution of large and small 
sunspot groups over four solar cycles. \apj\  731, 30.
 \doiurl{10.1088/0004-637X/731/1/30}.
\bibitem[\protect\citeauthoryear{Komitov}{2019}]{kom19}
Komitov, B., 2019. The 24$^{th}$ solar cycle: Preliminary analysis and 
generalizations. {\it BlgAJ} 30, 3.
\bibitem[\protect\citeauthoryear{Kumar \etal}{2022}]{kumar22}
Kumar, P., Biswas, A., Karak, B.B., 2022.  Physical link of the polar
 field buildup with the Waldmeier effect broadens the scope of early
 solar cycle prediction: Cycle 25 is likely to be slightly stronger than
 Cycle 24. \mnras\ 513, L112. \doiurl{10.1093/mnrasl/slac043}.
\bibitem[\protect\citeauthoryear{Kumar \etal}{2021}]{kumar21}
Kumar, P., Nagy, M., Lemerle, A., Karak, B.B., Petrovay, K. 2021. The polar 
precursor method for solar cycle prediction: Comparison of predictors and 
their temporal range. \apj\ 909, 87.  \doiurl{10.3847/1538-4357/abdbb4}.
\bibitem[\protect\citeauthoryear{Labonville \etal}{2019}]{lab19}
Labonville, F., Charbonneau, P., Lemerle, A., 2019. A Dynamo-based forecast
 of solar cycle 25. \solphys\  294, 82.
 \doiurl{10.1007/s11207-019-1480-0}.
\bibitem[\protect\citeauthoryear{Li \etal}{2018}]{li18}
Li, F.Y., Kong, D.F., Xie, J.L., Xiang, N.B., Xu, J.C., 2018.
 Solar cycle characteristics and their application in the prediction of 
cycle 25. {\it J. Atmos. Solar-Terr. Phys.} 181, 110.
  \doiurl{10.1016/j.jastp.2018.10.014}.
\bibitem[\protect\citeauthoryear{Lu \etal}{2022}]{lu22}
Lu, J.Y., Xiong, Y.T., Zhao, K., Wang, M., Li, J.Y., Peng, G.S., Sun, M., 2022.
 A Novel Bimodal Forecasting Model for Solar Cycle 25.\apj\ 924, 59.
 \doiurl{10.3847/1538-4357/ac3488}.
\bibitem[\protect\citeauthoryear{Luo and Tan}{2024}]{luo24}
Luo, P-X., Tan, B-L., 2024. Long-term evolution of solar activity and
 prediction of the following solar cycles. {\it Res. in Astron. and Astrophys.} 
24, 035016 (11pp). \doiurl{10.1088/1674-4527/ad1ed2}.
\bibitem[\protect\citeauthoryear{McCracken \etal}{2013}]{mcc13}
McCracken, K.G., Beer, J., Steinhilber, F., Abreu, J., 2013. A Phenomenological
study of the cosmic ray variations over the past
9400 years, and their implications regarding solar activity and
the solar dynamo. \solphys\ 286, 609.
 \doiurl{10.1007/s11207-013-0265-0}.
\bibitem[\protect\citeauthoryear{Mandal \etal}{2016}]{mandal16}
Mandal, S., Banerjee, D., 2016. Sunspot sizes and the solar
 cycle: Analysis using Kodaikanal white-light digitized data.
 \apjl\ 830, L33.  \doiurl{10.3847/2041-8205/830/2/L33}.
\bibitem[\protect\citeauthoryear{Nandy}{2021}]{nandy21}
Nandy, D., 2021. Progress in solar cycle predictions:
 Sunspot Cycles 24\,--\,25 in
 perspective, \solphys\ 296, 54. \doiurl{10.1007/s11207-021-01797-2}.
\bibitem[\protect\citeauthoryear{Nagovtsyn and Ivanov}{2023}]{nag23}
Nagovitsyn, Y.A., Ivanov, V.G., 2023. Solar cycle paring and prediction
 of cycle 25. \solphys\ 298, 37. \doiurl{10.1007/s11207-023-02121-w}.
\bibitem[\protect\citeauthoryear{Okoh \etal}{2018}]{okoh18}
Okoh, D.I., Seemala, G.K., Rabiu, A.B., Uwamahoro, J., Habarulema, J.B.,
 Aggarwal, M., 2018. A hybrid regression-neural network (HR-NN) method
 for forecasting the solar activity. {\it Space Weather} 16, 1424.
 \doiurl{10.1029/2018SW001907}.
\bibitem[\protect\citeauthoryear{Pesnell}{2012}]{pes12}
Pesnell, W.D., 2012. Solar cycle predictions (invited review).
 \solphys\ 281, 507.  \doiurl{10.1007/s11207-012-9997-5}.
\bibitem[\protect\citeauthoryear{Pesnell}{2018}]{pesnell18}
Pesnell, W.D., 2018. Effects of version 2 of the international sunspot number on
na\"ive predictions of solar cycle 25. {\it Space Weather}  16, 1997.
 \doiurl{10.1029/2018SW002080}.
\bibitem[\protect\citeauthoryear{Pesnell and Schatten}{2018}]{ps18}
Pesnell, W.D., Schatten, K.H., 2018. An early prediction of the 
amplitude of solar cycle 25. \solphys\ 293, 112.
 \doiurl{10.1007/s11207-018-1330-5}.
\bibitem[\protect\citeauthoryear{Petrovay}{2020}]{petro20}
Petrovay, K., 2020. Solar cycle prediction. 
{\it Liv. Rev. Solar Phys.} 17, 2. {10.1007/s41116-020-0022-z}.
\bibitem[\protect\citeauthoryear{Petrovay \etal}{2018}]{petro18}
Petrovay, K., Nagy, M., Gerj\`{a}k, T., Juh\`{a}sz, L., 2018. Precursors of
 an upcoming solar cycle at high latitudes from coronal green line data.
 {\it J. Atmos. Solar-Terr. Phys.} 176, 15.
 \doiurl{10.3103/S0884591308050036}.
\bibitem[\protect\citeauthoryear{Ramesh}{2000}]{ramesh00}
Ramesh, K.B., 2000. Dependence of $SSN_M$ on $SSN_m$ - a reconsideration for
 predicting the amplitude of a sunspot cycle. \solphys\ 197, 421.
 \doiurl{10.1023/A:1026565028898}.
\bibitem[\protect\citeauthoryear{Ramesh and Bhagya Lakshmi}{2012}]{ramesh12}
Ramesh, K.B., Bhagya Lakshmi, N., 2012. The amplitude of sunspot minimum as a
 favorable precursor for the prediction of the amplitude of the
next solar maximum and the limit of the Waldmeier effect.
 \solphys\ 276, 395.  {10.1007/s11207-011-9866-7}.
\bibitem[\protect\citeauthoryear{Ramesh and Rohini}{2008}]{ramesh08}
Ramesh, K.B., Rohini, V.S., 2008. 1.8 \o{A}, Coronal cackground X-ray
 emmission and the associated indicators of photospheric magnetic activity. 
 \apjl\  686, L41.  \doiurl{10.1086/592774}.
\bibitem[\protect\citeauthoryear{Singh and Bhargawa}{2017}]{sb17}
Singh, A.K., Bhargawa, A., 2017. An early prediction of 25th solar cycle 
using Hurst exponent. \apss\ 362, 199.
 \doiurl{10.1007/s10509-017-3180-2}.
\bibitem[\protect\citeauthoryear{Stefani \etal}{2021}]{stef21}
Stefani, F., Stepanov, R., Weler, T., 2021. Shaken and Stirred: When Bond meets 
Suess--de Vries and Gnevyshev--Ohl. \solphys\ 296, 88. 
 \doiurl{10.1007/s11207-021-01822-4}.
\bibitem[\protect\citeauthoryear{Tlatov}{2009}]{tlatov09}
Tlatov, A.G., 2009. The minimum activity epoch as a precursor of the
 solar activity. \solphys\ 260, 465. 
 \doiurl{10.1007/s11207-009-9451-5}.
\bibitem[\protect\citeauthoryear{Veronig \etal}{2021}]{veronig21}
Veronig, A.M., Jain, S., Podladchikova,T., P\"otzi, W., Clette, F., 2021 
Hemispheric sunspot numbers 1874\,--\,2020. \aap\ 652, 56.
 \doiurl{10.1051/0004-6361/202141195}.
\bibitem[\protect\citeauthoryear{Wilson}{2013}]{irgw13}
Wilson, I.R.G., 2013. The Venus--Earth--Jupiter spin--orbit coupling model.
{\it Pattern Recogn. Phys.} 1, 147. \doiurl{10.1086/508013}.
\bibitem[\protect\citeauthoryear{Wilson and Hathaway}{2006}]{wh06}
Wilson, R.M., Hathaway, D.H., 2006. On the relation between sunspot area and 
sunspot number. NASA/STI/Recon. Tech. Rep. No. 6.
\bibitem[\protect\citeauthoryear{Wood}{1972}]{wood72}
Wood, K., 1972. Sunspots and planets. \nat\, 240 (5376), 91. 
\doiurl{10.1038/240091a0}.
\bibitem[\protect\citeauthoryear{Wood and Wood}{1965}]{ww65}
Wood, R.M., Wood, K.D., 1965. Solar motion and sunspot comparison \nat\, 208,
 129. \doiurl{10.1038/208129a0}.
\bibitem[\protect\citeauthoryear{Zhu \etal}{2022}]{zhu22}
Zhu, H., Zhu, W., He, M., 2022. Prediction using an optimized long short-term 
memory mode with F10.7. \solphys\ 297, 157.
 \doiurl{10.1007/s11207-022-02091-5}.
\end{thebibliography}
\end{document}